\documentclass[sigconf]{acmart}
\AtBeginDocument{%
  }

\setcopyright{acmlicensed}
\copyrightyear{2025}
\acmYear{2025}
\acmDOI{10.1145/3746027.3754982}
\acmConference[MM '25] {Proceedings of the 33rd ACM International Conference on Multimedia}{October 27--31, 2025}{Dublin, Ireland.}
\acmBooktitle{Proceedings of the 33rd ACM International Conference on Multimedia (MM '25), October 27--31, 2025, Dublin, Ireland}
\acmISBN{979-8-4007-2035-2/2025/10}
\usepackage{multirow,subfigure,xcolor,balance}

\begin{document}

\title{Audio Does Matter: Importance-Aware Multi-Granularity Fusion for Video Moment Retrieval}

\author{Junan Lin}
\authornotemark[1]
\affiliation{%
  \institution{Zhejiang University}
  \city{Hangzhou}
  \country{China}
}
\email{linja@zju.edu.cn}

\author{Daizong Liu}
\authornote{Both authors contributed equally to this research.}
\affiliation{%
  \institution{Peking University}
  \city{Beijing}
  \country{China}}
\email{dzliu@stu.pku.edu.cn}

\author{Xianke Chen}
\affiliation{%
  \institution{Zhejiang Gongshang University}
  \city{Hangzhou}
  \country{China}
}
\email{a397283164@163.com}

\author{Xiaoye Qu}
\affiliation{%
  \institution{Shanghai Artificial Intelligence Laboratory}
  \city{Shanghai}
  \country{China}
}
\email{xiaoye@hust.edu.cn}

\author{Xun Yang}
\affiliation{%
  \institution{University of Science and Technology of China}
  \city{Hefei}
  \country{China}
}
\email{hfutyangxun@gmail.com}

\author{Jixiang Zhu}
\affiliation{%
  \institution{Zhejiang Gongshang University}
  \city{Hangzhou}
  \country{China}
}
\email{zhujx@mail.zjgsu.edu.cn}

\author{Sanyuan Zhang}
\affiliation{%
  \institution{Zhejiang University}
  \city{Hangzhou}
  \country{China}
}
\email{syzhang@zju.edu.cn}

\author{Jianfeng Dong}
\authornotemark[2]
\affiliation{%
  \institution{Zhejiang Gongshang University}
  \city{Hangzhou}
  \country{China}
}
\email{dongjf24@gmail.com}
\authornote{Corresponding author.}

\renewcommand{\shortauthors}{Junan Lin et al.}

\begin{abstract}
Video Moment Retrieval (VMR) aims to retrieve a specific moment semantically related to the given query. To tackle this task, most existing VMR methods solely focus on the visual and textual modalities while neglecting the complementary but important audio modality. Although a few recent works try to tackle the joint audio-vision-text reasoning, they treat all modalities equally and simply embed them without fine-grained interaction for moment retrieval. These designs are counter-practical as: Not all audios are helpful for video moment retrieval, and the audio of some videos may be complete noise or background sound that is meaningless to the moment determination. To this end, we propose a novel Importance-aware Multi-Granularity fusion model (IMG), which learns to dynamically and selectively aggregate the audio-vision-text contexts for VMR. Specifically, after integrating the textual guidance with vision and audio separately, we first design a pseudo-label-supervised audio importance predictor that predicts the importance score of the audio, and accordingly assigns weights to mitigate the interference caused by noisy audio. Then, we design a multi-granularity audio fusion module that adaptively fuses audio and visual modalities at local-, event-, and global-level, fully capturing their complementary contexts. We further propose a cross-modal knowledge distillation strategy to address the challenge of missing audio modality during inference. To evaluate our method, we further construct a new VMR dataset, \textit{i.e.}, Charades-AudioMatter, where audio-related samples are manually selected and re-organized from the original Charades-STA to validate the model's capability in utilizing audio modality. Extensive experiments validate the effectiveness of our method, achieving state-of-the-art with audio-video fusion in VMR methods. Our code is available at \url{https://github.com/HuiGuanLab/IMG}.
\end{abstract}





\maketitle

\section{Introduction}
\label{sec:intro}

Video Moment Retrieval (VMR) \cite{anne2017localizing,gao2017tall,yang2024learning,dong2024temporal,yang2024learning} aims to retrieve the part of the video that is relevant to the semantic of a given query. 
As a fundamental yet important task, it requires in-depth interaction between video and text semantics for accurate alignment and reasoning.
Existing mainstream works \cite{zhang2020span,zhang2020learning,zhang2021parallel,yang2022entity,zhang2023text,yang2024dynamic,dong2022dual,dong2023dual} generally focus on naive visual and textual modalities and develop vision-text integration frameworks to retrieve the specific moment. However, in addition to the visual contexts, audio modality also contains valuable contexts within the video streams \cite{bagchi2021hear,pan2022wnet,mercea2022temporal,jiang2022dhhn,mercea2022audio,zhou2021joint,gogate2020cochleanet}.
Without considering the rich complementary contexts of the audio modality, previous VMR methods fail to distinguish different activities like ``laughing" and ``talking" that share a similar visual appearance.
Therefore, exploring the interaction and fusion of audio, vision, and text modalities in VMR is a promising direction with research forward.

To leverage audio information, several audio-based VMR methods ~\cite{chen2023curriculum,liu2022umt,chen2020learning} have been proposed.
However, these approaches typically extract features from audio, vision, and text modalities and apply a uniform aggregation strategy for joint reasoning, without considering their diverse contributions.
For instance, PMI-LOC~\cite{chen2020learning} incorporates RGB, motion, and audio modalities, establishes pairwise interactions between modalities,
UMT \cite{liu2022umt} introduces a unified multimodal transformer framework for the integration of visual and audio information, while ADPN \cite{chen2023curriculum} leverages the consistency and complementarity between audio and visual modalities for efficient audio fusion.
Although these methods achieve relatively better performance than conventional VMR methods, they neglect
that not all audios contribute to the final grounding as the audio of some videos may be complete noise or background sound.

\begin{figure}[t!]
  \centering
   \includegraphics[width=0.9\linewidth]{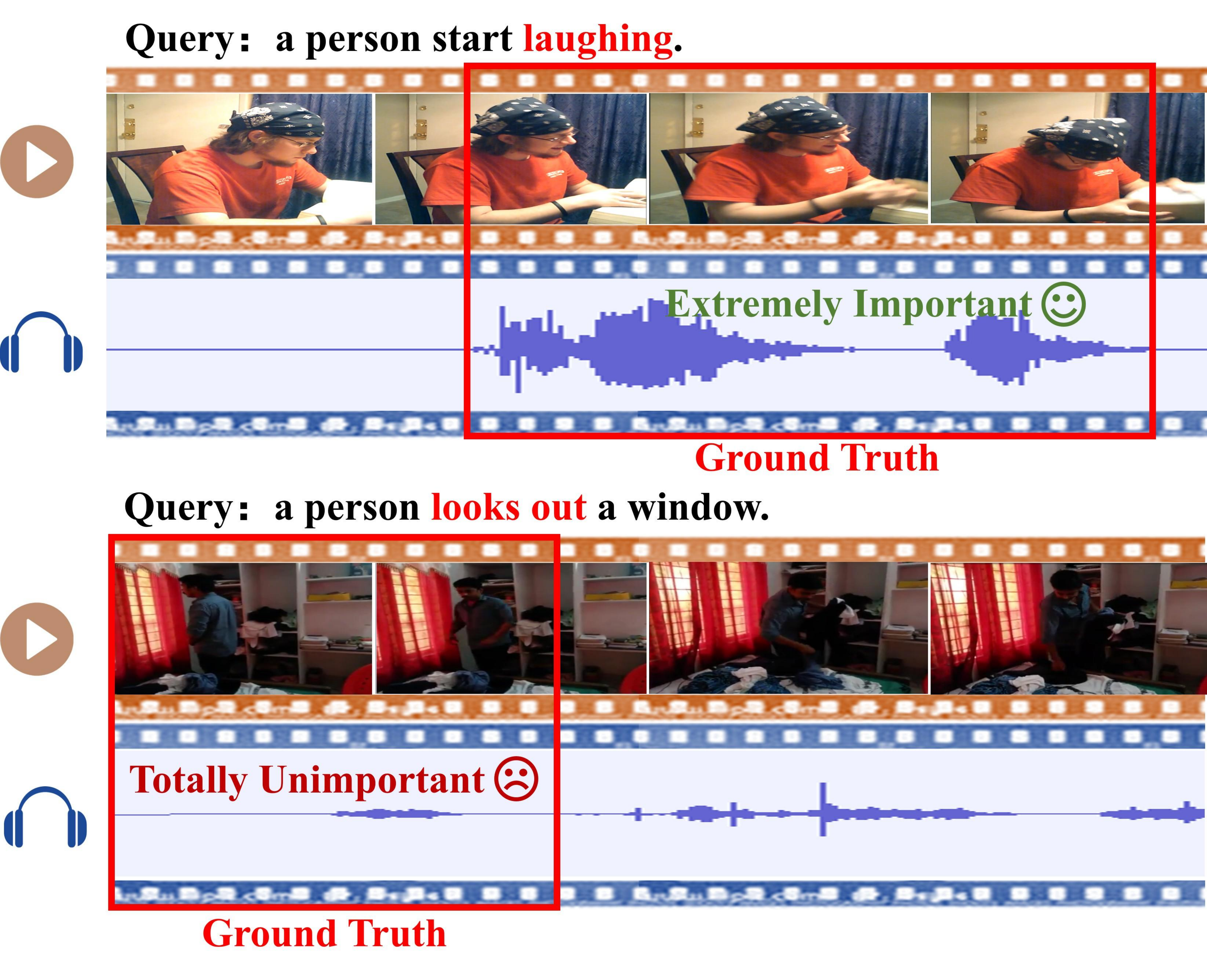} 
   \vspace{-3mm}
   \caption{
   (Top) Audio is a critical modality, outweighing the importance of vision. (Bottom) Audio is entirely irrelevant and considered noise relative to the vision.
   }
   \vspace{-6mm}
   \label{fig:intro}
\end{figure}

In practice, audio semantics exhibit considerable complexity and diversity, varying significantly across different scenarios. In certain instances, audio context serves as a valuable complement, enhancing the alignment with text semantics and facilitating accurate reasoning. Conversely, noisy audio can lead to erroneous textual associations. 
As shown in Figure~\ref{fig:intro}, for the first query ``a person is laughing", 
leveraging audio context significantly aids in identifying the laughing action, which might be ambiguous using vision alone.
However, for the second query ``a person looks out a window", 
the audio modality provides no benefit and may even be detrimental, given that this action is primarily visually driven.
Therefore, this motivates us to design a more dynamic audio-vision-text association framework that learns to selectively and adaptively aggregate appropriate contexts from audio and visual modalities for reasoning the specific text semantics.

To this end, we make the first attempt to tackle a flexible audio-vision-text joint reasoning for the VMR task. 
In particular, we propose a novel Importance-aware Multi-Granularity fusion model (IMG) with three prediction branches: audio branch, visual branch, and audio-visual fusion branch. 
Initially, textual guidance is integrated separately with both visual and audio inputs. We then introduce an audio importance-aware module to tackle the issue of variable audio importance, which is crucial in vision-text pairs. This module is supervised by pseudo-labels derived from the retrieval loss of each branch. It effectively learns to assess the relative importance of audio compared to vision. 
Then, for the latter audio-vision context fusion, we design a multi-granularity fusion network, which establishes local-level and event-level to global-level audio-vision fusion, as a way to better discover key clues in audio for assisting text-specific activity understanding within video contents. In addition to using traditional retrieval loss for supervision, since the multi-modal fusion branch tends to show better performance than the individual vision/audio reasoning branch as the former fuses the positive contexts from both modalities, we also distillate the knowledge from the fusion branch to the weaker visual and audio branches, thus strengthening the performance of both branches and in-turn providing better feedback to the fusion branch to further improving the performance. 

To sum up, the key contributions of our work are four-fold:
\begin{itemize}
    \item We propose a novel Importance-aware Multi-Granularity fusion network (IMG) to handle the audio-incorporated VMR task, which selectively fuses audio modal information of video samples at multiple granularities for final retrieval.
    \item We introduce an audio importance predictor, guided by a loss-aware pseudo-importance generator during training, to identify and emphasize semantically relevant audio clues. This mechanism enables the model to selectively focus on informative audio cues while suppressing irrelevant or noisy background sounds.  
    \item We propose a cross-modal knowledge distillation strategy, which transfers knowledge from the more effective fusion branch to the unimodal branch.  This strategy enables our framework to retain strong performance even when audio information is missing during inference.
    \item In addition to standard benchmarks such as Charades-STA and ActivityNet Captions, we introduce a new evaluation dataset, Charades-AudioMatter, where sample's audio matter for moment retrieval. Extensive experiments on these datasets demonstrate the effectiveness of our approach, particularly in scenarios where audio cues play a complementary or dominant role.
\end{itemize}

\section{Related Works}
\label{sec:relatework}
\noindent \textbf{Video Moment Retrieval (VMR)}. VMR aims to retrieve a specific video segment based on a natural language query. Current approaches fall into two categories: proposal-based and proposal-free. For proposal-based ~\cite{zhang2020learning,sun2022you,shen2023semantics,wang2022negative,xiao2021boundary,zhang2021multi,liu2021adaptive,zheng2023progressive}, it is often necessary to pre-segment the candidate proposals, and the pre-segmented proposals and text are used as inputs to the cross-modal matching module for retrieval. For proposal-free ~\cite{zhang2020span,zhang2021parallel,huang2022video,yang2024dynamic,yang2022entity,zhang2023text,dong2024temporal,zhang2025video}, they eliminate the need for predefined proposals, processing raw visual and textual features directly through cross-modal matching. Building on these paradigms, recent studies~\cite{wang2023ms,moon2023query,jang2023knowing,sun2024tr,lee2024bam} have explored DETR-style architectures~\cite{carion2020end} to formulate VMR as a set prediction problem, enabling more flexible and end-to-end training.
Further extending these trends, some works aim to unify various video tasks (e.g., moment retrieval, highlight detection, video summarization) under a general framework~\cite{lin2023univtg,yan2023unloc}.
Meanwhile, the rapid progress of large language models (LLMs) has inspired a new wave of research that leverages their semantic reasoning capabilities to enhance VMR~\cite{wang2024hawkeye,ren2024timechat,wang2024grounded,guo2024vtg,jiang2024prior,yang2024robust}. Concurrently, audio has emerged as a valuable modality for complementing vision in VMR, \textit{e.g.}, PMI-LOC \cite{chen2020learning} employs RGB, motion, and audio and is designed to interact with pairs of modalities at the sequence and channel levels. UMT \cite{liu2022umt} proposes a unified multimodal transformer framework to fuse vision and audio. ADPN \cite{chen2023curriculum} proposes a text-guided clues miner to fill the information gap between audio-visual modalities. However, the above models overlook the inherent uncertainty of audio as a modality and the contribution of audio varies significantly depending on the specific query and video content, highlighting a need for more adaptive solutions.

\begin{figure*}[t]
  \centering
   \includegraphics[width=1\linewidth]{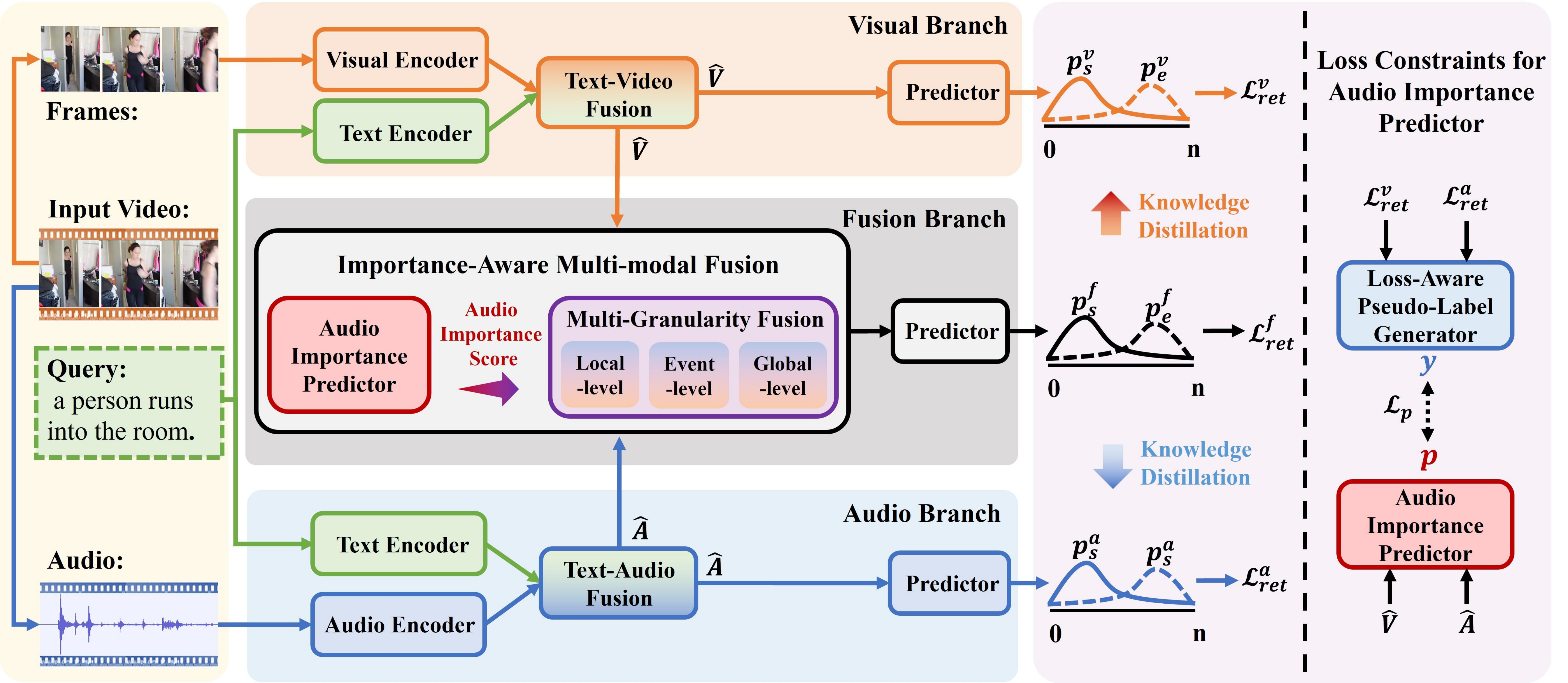} 
   \caption{
   The framework of our proposed importance-aware multi-granularity fusion model for video moment retrieval.
   }
   \label{fig:framework}
\end{figure*}

\noindent \textbf{Uncertain Modal Learning.}  The audio modality often exhibits uncertainty and imbalance in video comprehension tasks \cite{xu2023mmcosine,fu2023multimodal}. For instance, audio in some videos may consist solely of noise or background sounds, while in text-based video tasks, the query may be entirely independent of the audio. Similar issues arise in other modalities and these modal imbalance challenges have garnered significant attention \cite{lin2024multi,ren2024novel,yao2024drfuse,huang2021learning,xu2022different,cheng2019noise}. To address these challenges, Li \textit{et al.} \cite{li2024prototype} quantified uncertainty caused by inherent data ambiguity to enhance prediction reliability. Tellamekala \textit{et al.} \cite{tellamekala2023cold} addressed modal uncertainty in categorical sentiment recognition by introducing a modeling approach that enforces both calibration and ordinality constraints and Zhang \textit{et al.} \cite{zhang2024multimodal} explored the challenges and solutions for low-quality multimodal fusion, emphasizing the promise of dynamic multimodal learning in overcoming sample-specific, temporal, and spatial variations.
Building on these developments, we introduce the audio importance predictor. Supervised by dynamic pseudo-labels derived from sample-wise loss functions, this predictor quantifies the audio modality's importance, providing a critical parameter for adaptive modal fusion.

\section{Method}
\label{sec:method}

\subsection{Overview}
\noindent \textbf{Problem Definition.}
Video moment retrieval aims to retrieve the start-end frame pair $\{f_s,f_e\}$ of a specific segment that semantically match the textual query $Q=\{w_i\}_{i=1}^{N}$ from the untrimmed video $V=\{f_t\}_{t=1}^{T}$, where $w_i$ represents the $i$-th word and $f_t$ represents the $t$-th frame.
Additionally, for each video frame, we can extract an audio-aware clip as a complementary modality. Thus, the corresponding audio stream is represented as $A=\{a_j\}_{j=1}^{T}$, where $a_j$ denotes the $j$-th audio clip, providing contextual knowledge to enhance the retrieval process.

\noindent \textbf{Overall Pipline.} We illustrate our proposed framework in Figure~\ref{fig:framework}.
Given the pre-extracted visual, audio, and text features by the corresponding encoder, our IMG model first employs feed-forward network (FFN) layers to map these features into a common latent space. Then, we employ interactions between vision-text and audio-text pairs, fusing them to derive text-semantic-activated visual and audio features. These features are then passed into the visual-audio fusion branch, where they dynamically interact to enable joint reasoning. Specifically, an audio importance predictor generates a sample-wise score, which serves as a crucial weight to determine the audio-to-vision complementarity coefficients for the given sample pair.    
Then, the visual and audio features will be fed into a multi-granularity fusion module to aggregate the target-moment-related information at local-, event-, and global-levels according to the previously obtained important weight.
Finally, the three-level features are concatenated and fed into the predictor to output predictions, while the visual-only feature and audio-only feature are also fed into their respective unimodal predictors. A multi-branch training with cross-modal knowledge distillation strategy is used to transfer knowledge from the fusion branch to the unimodal branch. During inference, the retrieval branch can be freely selected, with the fusion branch typically being the preferred choice.

\subsection{Input Representation}
\label{sec:fe}
\noindent \textbf{Multi-Modal Feature Representation.} \textit{For audio} modality, we firstly use pre-trained audio-aware CNN \cite{kong2020panns,hershey2017cnn} to extract its original features $A\in\mathbb{R}^{{T}\times{d_a}}$, then employ an audio encoder which is composed by an FFN, convolutional and transformer layers on them follow \cite{zhang2020span}, textual dependency enhanced features $A'\in\mathbb{R}^{{T}\times{d}}$.
\textit{For vision} modality, we extract the original visual features $V\in\mathbb{R}^{{T}\times{d_v}}$ by a pre-trained visual CNN \cite{carreira2017quo,tran2015learning,zhou2025egotextvqa} and further obtain corresponding enhanced features  $V'\in\mathbb{R}^{{T}\times{d}}$ by a visual encoder which shares the same structure as audio encoder.
\textit{For textual query}, we directly initialize it with GloVe embeddings \cite{pennington2014glove}. Since the query may have different semantic alignments with vision and audio, 
we further encode it by two separate text encoders which also share the same structure as the audio encoder and obtain modal-specific enhanced text features as $Q'_a\in\mathbb{R}^{{N}\times{d}}$ and $Q'_v\in\mathbb{R}^{{N}\times{d}}$. 


\noindent \textbf{Vision-text/Audio-Text Fusion.}
To highlight the most related contents between the vision/audio and the given textual query, we apply context-query attention \cite{zhang2020span} on each pair, resulting in fused features $\hat{V}\in\mathbb{R}^{{T}\times{d}}$  and $\hat{A}\in\mathbb{R}^{{T}\times{d}}$.

\subsection{Importance-Aware Multi-modal Fusion}
The importance-aware multi-modal fusion is a multi-granularity fusion module guided by an audio importance predictor. The predictor is trained to identify and emphasize semantically relevant audio cues, enabling the model to selectively fuse informative audio signals with visual features. Guided by the predicted importance scores, the multi-granularity fusion process effectively filters out irrelevant or noisy audio content while aggregating meaningful cross-modal information at multiple temporal levels, thereby enhancing retrieval performance.

\label{sec:IAMF}
\subsubsection{Audio Importance Predictor} 
\label{sec:AIP}
The Audio Importance Predictor (AIP) is a lightweight module designed to dynamically estimate the relative importance of audio for each video-query pair. Since ground-truth importance labels are unavailable, we design a loss-aware pseudo-importance generator to produce pseudo labels that serve as supervision signals during training.


\noindent \textbf{Structure}.
Given the text-guided visual and audio features $\hat{V}$ and $\hat{A}$, we first apply attention pooling \cite{bahdanau2014neural} to obtain their global representations, denoted as $\hat{V}_G \in \mathbb{R}^{d}$ and $\hat{A}_G \in \mathbb{R}^{d}$. These global features capture the overall semantic context of the visual and audio modalities, respectively.
Next, we concatenate the two global features and feed them into a Multi-Layer Perceptron (MLP), which facilitates mutual feature interaction and enables the model to reason about the relative importance of audio with respect to the visual context. The audio importance score $p$ is then predicted as:
$p=Sigmoid(MLP([\hat{A_G};\hat{V_G}]))$, 
where $Sigmoid$ denotes the sigmoid activation function, and $[;]$ indicate the concatenation operator.
This predicted score $p$ serves as a sample-wise importance weight, guiding the subsequent multimodal fusion by modulating the contribution of the audio modality.

\noindent \textbf{Training with the pseudo importance labels.} 
In order to train the audio importance predictor, we should construct pseudo labels as supervisory signals.
We draw inspiration from the observation that neural networks tend to prioritize learning from simpler samples, which typically correspond to lower training losses \cite{arpit2017closer}. Based on this, we compare the retrieval losses of the audio and visual branches for each video-query pair. The modality with a lower loss is considered to provide more relevant information and is thus assigned a higher pseudo-importance score.
Specifically, we compute a pseudo-importance score $y'$ with a softmax-like normalization:
\begin{equation}
 y=\frac{e^{\mathcal{L}^v_{ret}/\gamma}}
{e^{\mathcal{L}^a_{ret}/\gamma}+e^{\mathcal{L}^v_{ret}/\gamma}}, 
 \\
 y'=
\begin{cases}
1 & \text{if } y \geq \epsilon_{max}, \\
y & \text{if } \epsilon_{max} > y \geq \epsilon_{min} \\
0 & \text{if } y < \epsilon_{min},
\end{cases},
\end{equation}
where $\mathcal{L}^a_{ret}$ and $\mathcal{L}^v_{ret}$ represents the retrieval loss of audio branch and visual branch, respectively, $\gamma$ is temperature coefficient.
Besides, $\epsilon_{min}$ is a lower threshold below which the audio modality is considered uninformative, and its contribution is suppressed. Conversely, values above $\epsilon_{max}$ indicate that audio plays a dominant role in retrieval.
Finally, we use a binary cross entropy loss to train AIP as:
\begin{equation}
\mathcal{L}_{p}=\frac{1}{B}\sum_{i=1}^{B} y'_i log p_i + (1-y'_i)log(1-p_i), 
\end{equation}
where $B$ denotes batch size, and $i$ represents the index of $i$-th sample.

The predicted importance score $p$ serves as a key control parameter in the subsequent multi-granularity fusion stage, guiding the selective integration of audio and visual features.
To prevent unstable predictions from misguiding early fusion, we initialize the fusion weight with a neutral value of 0.5 and gradually increase the influence of the AIP-predicted score as training progresses. This curriculum-like strategy helps the model build robust multimodal interactions while mitigating the impact of early-stage noise in the importance estimation.


\begin{figure*}[t]
  \centering
   \includegraphics[width=0.8\linewidth]{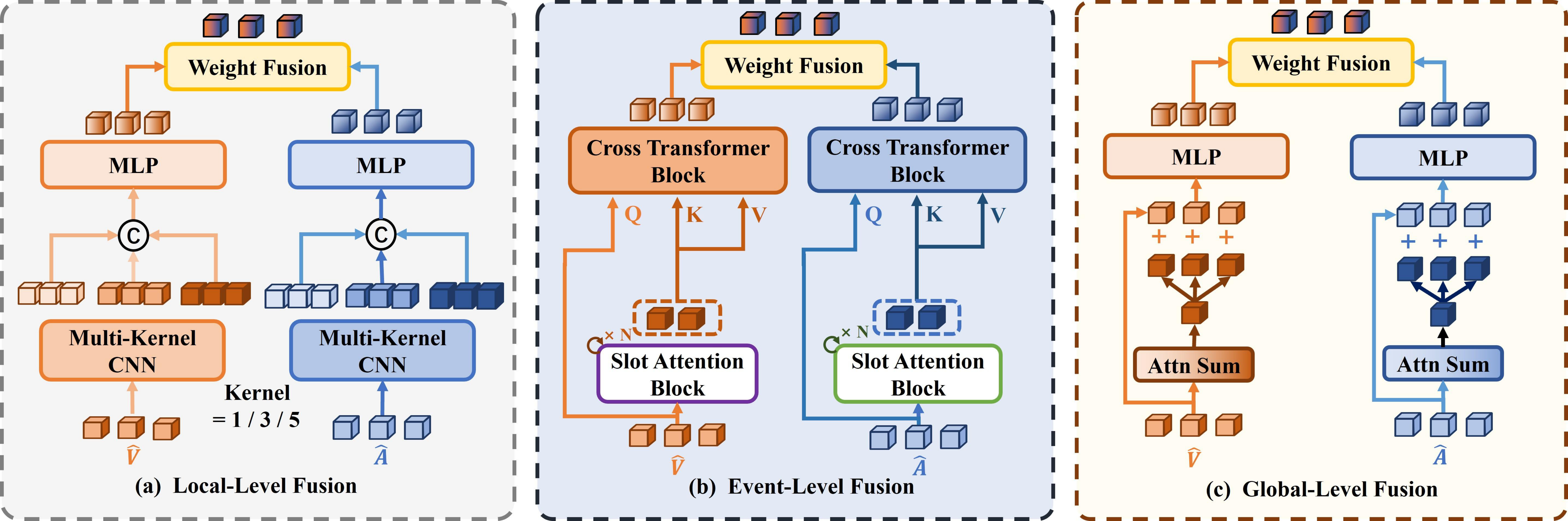}
   \caption{Our proposed Multi-Granularity Fusion module: (a) Local-Level Fusion, (b) Event-Level Fusion, (c) Global-Level Fusion.}
   \label{fig:lla&ela&gla}
\end{figure*}

\subsubsection{Multi-Granularity Fusion}
\label{sec:MGAM}

Given the inherently noisy and variable nature of the audio modality compared to visual signals, a simple fusion strategy may not be sufficient to fully exploit audio-visual complementarity. To address this, we propose a Multi-Granularity Fusion (MGF) module that performs hierarchical fusion from Local-, Event- and Global-perspective, and guided by the dynamically estimated audio importance score.

\noindent \textbf{Local-Level Visual-Audio Fusion.} As shown in Figure~\ref{fig:lla&ela&gla}(a), to match the visual and audio context frame-to-clip for fine-level fusion, we construct symmetric multi-kernel 1D convolutional networks and to deeply perceive the local relationships between video frames and audio clips, as follows:
\begin{equation}
\begin{aligned}
c^v_{k}=Conv1d_k(\hat{V}), \ c^a_{k}=Conv1d_k(\hat{A}), 
\end{aligned}
\end{equation}
where $k$ is the kernel size of the convolutional networks. The outputs are then concatenated and encoded by an MLP layer to map the dimensionality to $d$, as follows:
\begin{equation}
\begin{aligned}
\hat{V_l}=MLP([c^v_{1};...;c^v_{n}]), \hat{A_l}=MLP([c^a_{1};...;c^a_{n}]), 
\end{aligned}
\end{equation}
where $\hat{V_l} \in\mathbb{R}^{{T}\times{d}}$ and $\hat{A_l} \in\mathbb{R}^{{T}\times{d}}$. From this, we obtain the audio and visual features after reinforcement at the local level.

Finally, we fuse two features by element-wise addition with weight $p$ that is derived from the audio importance predictor:
\begin{equation}
\mathcal{F}_l=(1-p)LN(\hat{V_l})+pLN(\hat{A_l}),
\label{eq:fuse}
\end{equation}
where $LN$(·) is layer normalization.

\noindent \textbf{Event-Level Visual-Audio Fusion.}
As shown in Figure~\ref{fig:lla&ela&gla}(b), to match the event-aware semantics between vision and audio for activity reasoning, our event-level fusion module first employs a group of slot attention mechanism \cite{locatello2020object} to aggregate similar visual/audio clips into multiple events by using a set of learnable event slots, as follows:
\begin{equation}
\hat{A}_{s}=SlotAttn(\hat{A}),\hat{V}_{s}=SlotAttn(\hat{V}), 
\end{equation}
where $\hat{A}_{s}\in\mathbb{R}^{{e}\times{d}}$ and $\hat{V}_{s}\in\mathbb{R}^{{e}\times{d}}$ indicates that $e$ events are extracted from the visual/audio sequence. Subsequently, origin visual/audio features enter the cross-modal transformer layer as query and visual/audio events as key and value to obtain $\hat{A_e}$ and $\hat{V_e}$. Finally, we fuse to obtain visual-audio event aware features $\mathcal{F}_e$ as same as Equation ~\ref{eq:fuse}.

\noindent \textbf{Global-Level Visual-Audio Fusion.} As shown in Figure~\ref{fig:lla&ela&gla}(c), to match the visual and audio context from a global perspective, we first encode visual/audio features $\hat{V}$ and $\hat{A}$ into global level representation with attention pooling mechanism \cite{bahdanau2014neural}. Then, we concatenate it with each element of origin $\hat{V}$ and $\hat{A}$, and an MLP layer is used to obtain $\hat{V}_g$ and $\hat{A}_g$. Finally, we obtain visual-audio global aware features $\mathcal{F}_g$ as same as Equation ~\ref{eq:fuse}.

\noindent \textbf{Multi-Scale Feature Fusion.} Since the fused features obtained from different granularities have varying interrelationships, we adopt a set of Bi-GRUs to re-establish these inter-perceptual relationships by combining the features pairwise. Finally, we concatenate the results and pass them through MLP layers to map the dimensions back to $d$-dimension space, obtaining our final visual-audio fused features $\mathcal{F}$.

\subsection{Cross-modal Knowledge Distillation}


The fusion branch, by jointly modeling audio and visual cues, inherently captures richer and more comprehensive semantic representations. However, in practical applications, audio signals may be missing, corrupted, or unavailable during inference. To ensure that unimodal branches retain strong retrieval capabilities under such conditions, particularly for the visual branch, we introduce a cross-modal knowledge distillation strategy. Specifically, we treat the fusion branch as a teacher to distill knowledge into the unimodal branches, particularly the visual branch. This enables the unimodal branch to inherit modality-complementary cues from the fusion branch, thereby achieving strong retrieval performance even with visual-only input. To this end, we minimize the Kullback-Leibler (KL) divergence between the output distributions of the fusion and unimodal branches as follows:
\begin{equation}
\begin{aligned}
\mathcal{L}_{kl} = \sum_{i=1}^{B}\tau^2(KL(\sigma(s^s/\tau),\sigma(t^s/\tau)) \\
+KL(\sigma(s^e/\tau),\sigma(t^e/\tau))),
\end{aligned}
\end{equation}
where $s^{s/e}$ is start or end logits predicted by the student unimodal branch, $t^{s/e}$ is start or end logits predicted by the teacher fusion branch, $\tau$ is temperature coefficient and $\sigma$ is softmax function. Combining the two unimodal branches, the final KL divergence loss is the summation of the corresponding losses on the visual $\mathcal{L}^v_{kl}$ and the audio $\mathcal{L}^a_{kl}$.

\subsection{Model Training}
\label{sec:remain}

Following previous works \cite{zhang2020span}, we exploit the moment predictor as the retrieval heads to output the start logits and end logits of the moment, and get the final prediction $P_s$ and $P_e$.
Take the fusion branch as an example, the retrieval loss is computed as follows:
\begin{equation}
\mathcal{L}^f_{ret} = CE(P^f_s,Y_s) + CE(P^f_e,Y_e),
\end{equation}
where CE denotes cross-entropy loss, $Y_{s/e}=\{Y^i_{s/e}\}_i \in \{0,1\}$ represents the supervision where $Y_{s/e}$ is set to 1 only at the start/end point. By applying this loss function to the three branches (visual branch, audio branch and visual-audio fusion branch), the total retrieval loss of prediction is:
\begin{equation}
\mathcal{L}_{ret} = \mathcal{L}^v_{ret} + \mathcal{L}^a_{ret} + \mathcal{L}^f_{ret}.
\end{equation}

In addition, following \cite{lei2021detecting}, we also introduce saliency loss $\mathcal{L}_{sal}$ to the vision-text fusion features $\hat{V}$ and audio-text fusion features $\hat{A}$ as well as visual-audio fused features $\mathcal{F}$. This loss widens the distance between features within and outside of the timestamp. 
Finally, the overall training loss is:
\begin{equation}
\mathcal{L} = \mathcal{L}_{ret} + \lambda_{1}\mathcal{L}_{p} + \lambda_{2}\mathcal{L}_{kl} + \lambda_{3}\mathcal{L}_{sal},
\end{equation}
where $\lambda_{1}$, $\lambda_{2}$ and $\lambda_{3}$ are the balancing parameters.
During inference, we use Maximum Likelihood Estimation (MLE) to obtain the predicted $(y^s,y^e)$ with the constraint $y^s\leq y^e$.

\vspace{-8px}
\section{Experiment}
\subsection{Dataset}
We conduct our experiments on two video
moment retrieval benchmark datasets with audio, \textit{i}.\textit{e}., Charades-STA~\cite{gao2017tall}  and ActivityNet Captions \cite{krishna2017dense}, as well as Chardes-AudioMatter dataset reconstructed by ours. Specifically, \textbf{Charades-STA} is a dataset about daily indoor activities. There are 12,408 and 3,720 moment annotations for training and testing, respectively. \textbf{ActivityNet Captions} dataset contains about 20k videos taken from ActivityNet. We follow the setup in \cite{zhang2020span} with 37,421 moment annotations for training, and 17,505 annotations for testing. 

Moreover, to further validate the model's capability in integrating audio modality, we introduce a new dataset, \textbf{Charades-AudioMatter}, where audio matters for each test query. By reviewing both the videos and their corresponding audio, we manually select and re-organize 1,196 samples from the test set of Charades-STA in which the audio provides valuable information. This selection constitutes a new test set, while the training set of Charades-STA remained unchanged, see supplementary material for more details.

\subsection{Evaluation Metrics} 
Following the previous works \cite{gao2017tall,liu2018attentive,yuan2019find}, we adopt “R$n$@$\mu$” and “mIoU” as the evaluation metrics. The “R$n$@$\mu$” denotes the percentage of language queries having at least one result whose Intersection over Union (IoU) with ground truth is larger than $\mu*0.1$ in top-$n$ retrieved moments. “mIoU” is the average IoU over all testing samples. In our experiments, we use $n = 1$ and $ \mu\in \{3, 5, 7\}$.

\begin{table} [t!]
\caption{Ablation studies of Audio Importance Predictor (AIP) on Charades-STA.}
\label{tab:ab_mo}
\renewcommand{\arraystretch}{1.2}
\centering 
\scalebox{0.8}{
\begin{tabular}{cccccc}
\toprule
\textbf{Line ID} & \textbf{Approach} & \textbf{R1@3} & \textbf{R1@5} & \textbf{R1@7} & \textbf{mIOU} \\
\hline
\#1  & Add & 74.19 & 60.97 & 43.41 & 55.02
\\
\#2  & Concat & 72.77 & 59.73 & 43.20 & 54.24
\\
\#3  & Sim  & 74.33 & 60.91 & 43.80 & 55.12
\\
\#4  & Attn Entropy  & 73.76 & 60.11 & 43.23 & 55.00
\\
\#5 & AIP w/o pseudo-label & 73.98 & 59.74 & 43.33 & 54.39
\\
\hline
\#6 & AIP & \textbf{75.18} & \textbf{61.85} & \textbf{44.23} & \textbf{55.62}
\\
\bottomrule
\end{tabular}
}
\end{table}

\subsection{Ablation Study}
\noindent \textbf{Effectiveness of Audio Importance Predictor.} To demonstrate the effectiveness of our audio importance predictor, we conducted ablation studies. In our initial design, we explored multiple approaches for fusing audio and visual modalities. As shown in Table \ref{tab:ab_mo}, we compared various approaches with our weighted fusion based on predicted importance of AIP, including direct addtiton (line 1), concatenation (line 2), and weighted fusion based on cosine similarity between embeddings (line 3) and attention entropy (line 4) calculated by the last attention layer, we also compared AIP without the supervision of pseudo-label (line 5). Ultimately, our AIP demonstrated superior performance, which we attribute to our carefully designed label-supervised module that provides effective guidance for  AIP and ultimately achieve better dynamic multimodal integration.

\begin{figure}[t!]
  \centering
    \subfigure[Performance curves when noisy audio is introduced.]{\includegraphics[width=0.485\linewidth]{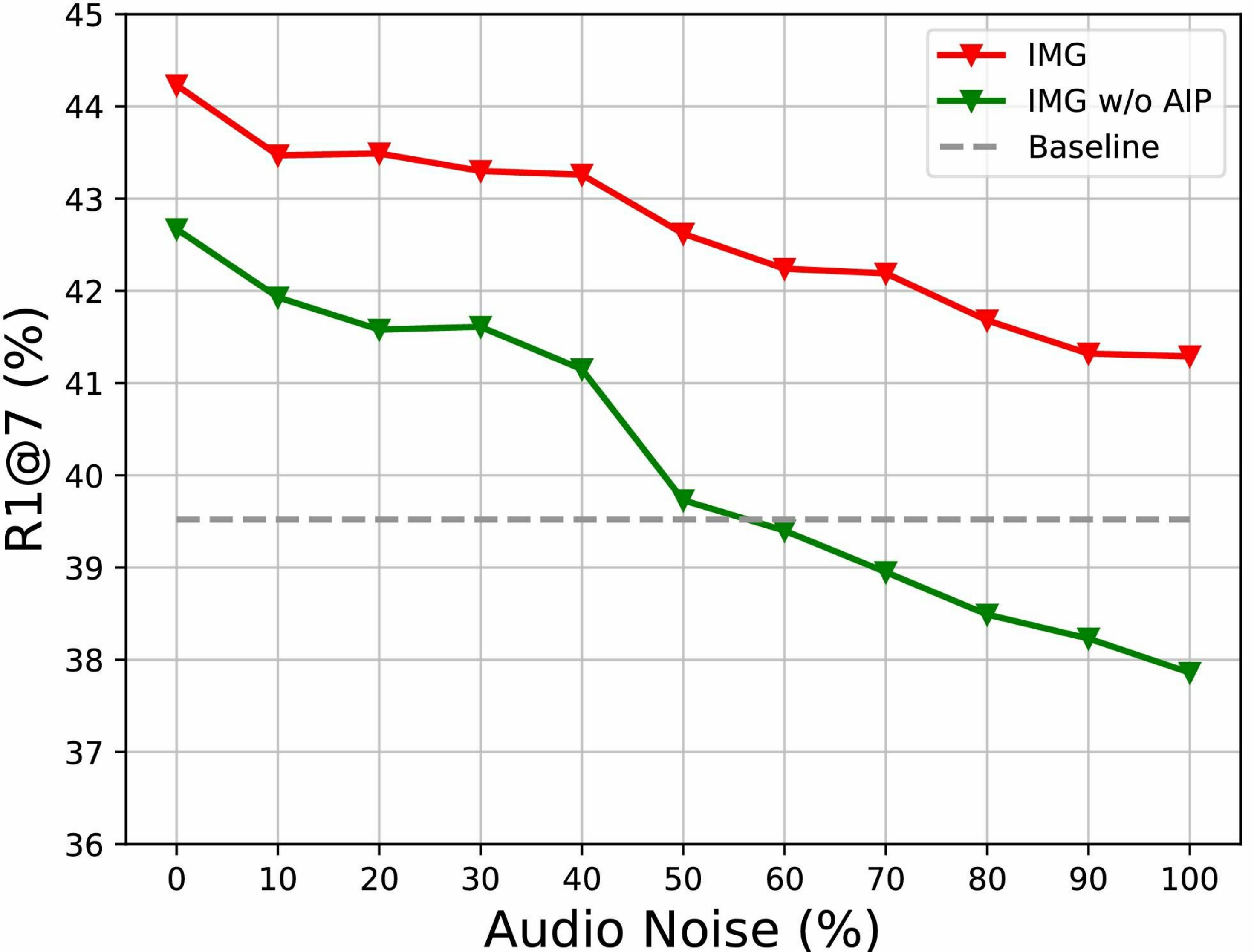}}
    \subfigure[Average audio importance curve predicted by AIP.]{\includegraphics[width=0.48\linewidth]{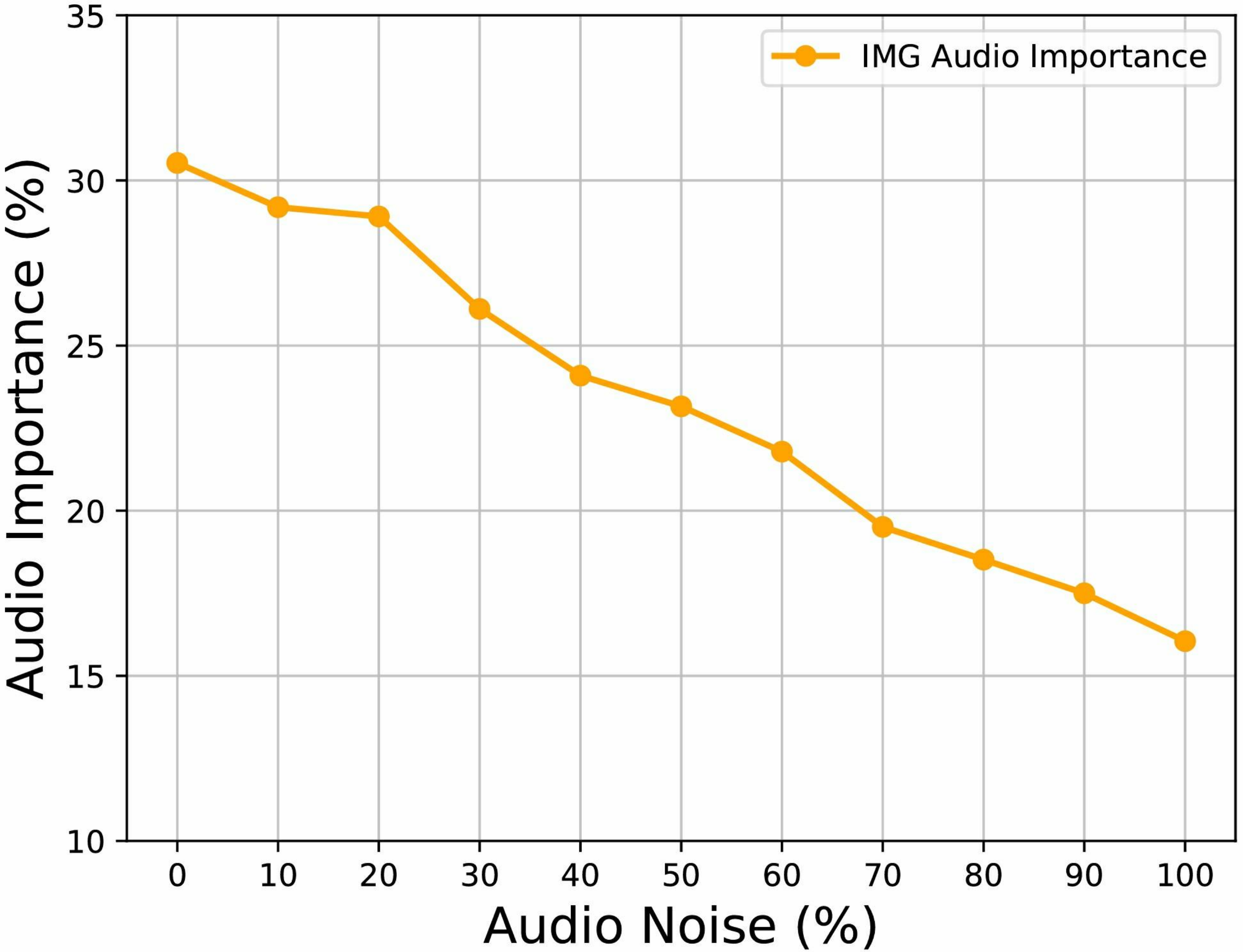}}
    \vspace{-10pt}
  \caption{During inference, as noise in the audio progressively increases, the gap between the two curves in (a) widens, suggesting that the IMG model with AIP exhibits greater robustness. Additionally, as we expected, the average audio importance in (b) decreases as noise levels rise.}
  \label{fig:ab_AIP}
\end{figure}

\noindent \textbf{Robustness Analysis on Audio Importance Predictor.} To assess the robustness of AIP, we introduced random gaussian noise to a subset of the test set audio samples. As shown in Figure \ref{fig:ab_AIP}, increasing the proportion of noisy audio widened the performance gap between IMG models with and without AIP. Notably, the performance of IMG without AIP fell below the baseline, while IMG with AIP exhibited a more gentle decline, proving the AIP's robustness and highlighting the detrimental impact on performance when modality importance is ignored under extreme modal imbalance.

\noindent \textbf{Effectiveness of Fusion Strategies.} 
We conduct ablation studies to evaluate the visual-audio fusion strategy in Table \ref{tab:ab_fuse}. Here, each of our proposed fusion methods demonstrates a performance improvement, highlighting the effectiveness of our fusion strategy. Performance is further enhanced when two feature aspects are fused, with the fusion of three feature sets outperforming the fusion of two. These results indicate that features extracted at different granularities complement each other effectively, facilitating a more comprehensive fusion of audio information.

\begin{table} [t!]
\caption{Ablation studies of fusion strategy on Charades-STA.}
\vspace{-4pt}
\label{tab:ab_fuse}
\renewcommand{\arraystretch}{1.2}
\centering 
\scalebox{0.8}{
\begin{tabular}{ccccccc}
\toprule
 \textbf{Local} & \textbf{Event} & \textbf{Global} & \textbf{R1@3} & \textbf{R1@5} & \textbf{R1@7} & \textbf{mIOU} \\
\hline
 \checkmark & - & - & 73.07 & 58.85 & 40.68 & 53.67 \\
 - & \checkmark & - & 74.84 & 59.92 & 41.32 & 54.83 \\
 - & - & \checkmark & 73.20 & 57.50 & 41.64 & 53.67 \\
\hline
 \checkmark & \checkmark & - & 74.09 & 60.08 & 42.64 & 55.08 \\
 \checkmark & - & \checkmark & 73.88 & 59.98 & 43.15 & 54.87 \\
 - & \checkmark & \checkmark & 74.28 & 60.33 & 42.93 & 55.47 \\
\hline
 \checkmark & \checkmark & \checkmark & \textbf{75.18} & \textbf{61.85} & \textbf{44.23} & \textbf{55.62}\\

\bottomrule
\end{tabular}
}
\end{table}
\begin{figure}[tb!]
  \centering
   \includegraphics[width=0.75\linewidth]{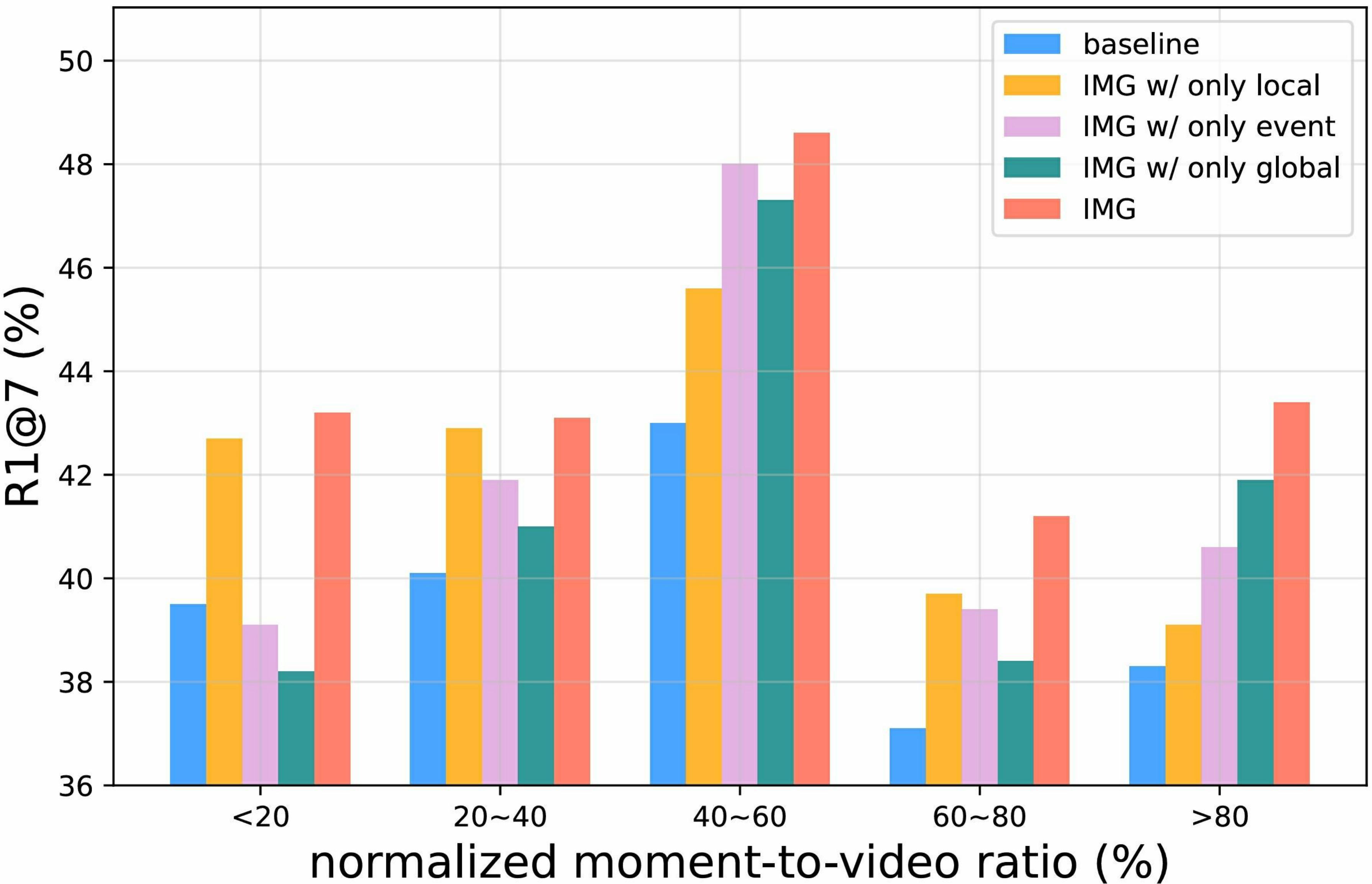}
   \vspace{-5pt}
   \caption{Performance of different granularity fusion strategies at different normalized moment-to-video ratios. 
   }
   \vspace{-10pt}
   \label{fig:mgf}
\end{figure}
\begin{table} [tb!]
\caption{Ablation studies on Charades-STA under conditions of using unimodal branch during inference.
“CKD” denotes Cross-modal Knowledge Distillation.
}
\label{tab:ab_noa}
\renewcommand{\arraystretch}{1.2}
\centering 
\scalebox{0.8}{
\begin{tabular}{lcllll}
\toprule
\textbf{Method} & \textbf{Branch} & \textbf{R1@3} & \textbf{R1@5} & \textbf{R1@7} & \textbf{mIOU} \\
\hline
\multirow{3}{*}{IMG} & Fusion & 75.18 & 61.85 & 44.23 & 55.62
\\
& Visual &  $74.84_{0.34\downarrow}$	& $60.95_{0.90\downarrow}$	& $43.44_{0.79\downarrow}$	& $54.97_{0.65\downarrow}$
\\
& Audio & $60.11_{15.07\downarrow}$	& $45.86_{15.99\downarrow}$	& $29.35_{14.88\downarrow}$	& $42.85_{12.77\downarrow}$
\\
\hline
\multirow{3}{*}{IMG w/o CKD} & Fusion & 74.09 & 61.03 & 43.33 & 55.31
\\
& Visual & $72.49_{1.60\downarrow}$ & $56.92_{4.11\downarrow}$ & $39.58_{3.75\downarrow}$ & $53.12_{2.19\downarrow}$
\\
& Audio & $58.04_{16.05\downarrow}$ & $43.70_{17.33\downarrow}$ & $25.48_{17.85\downarrow}$ & $40.68_{14.63\downarrow}$
\\
\bottomrule
\end{tabular}
}
\vspace{-6pt}
\end{table}

\noindent \textbf{Qualitative Analysis of Fusion Strategies.}This experiment will attempt to answer why our visual-audio fusion strategies is effective and can complement each other. The starting point of this structure is that we want different fusion strategies to focus on different sides of information, \textit{e.g.}, for local-level, our expectation is to find subtle clues. In Figure \ref{fig:mgf}, we categorized samples of Charades-STA into 5 equal number of categories based on the moment-to-video ratios, and we find that IMG with only local fusion show a stronger ability to handle smaller ratio (\textit{i.e.}, more subtle moments), whereas IMG with only event fusion enhanced performance of moderate ratio, and for IMG with only global fusion are suited to deal with the case of larger ration. The different performances between  different granularities lay the foundation for multi-granularity fusion and finally, our IMG achieves a good balance.

\noindent \textbf{Inference with Unimodal Branch.} In real-world scenarios, the audio modality may sometimes be irrelevant or unavailable, such as in surveillance footage. In such cases, the visual branch of IMG can still be employed, and cross-modal knowledge distillation strategy is expected to mitigate potential negative impacts. As shown in Table \ref{tab:ab_noa}, IMG with CKD exhibits minimal performance degradation which confirmed that CKD can largely overcome the negative effects and IMG still demonstrates excellent performance when audio modality is missing during inference. Additionally, we extend our investigation to the inference of audio branch, it turns out that audio branch inference alone has a significant degradation in performance, so we argue that audio can only be an auxiliary modality, and comparing the models with and without the CKD, we find that CKD also improves the performance of audio branch.

\noindent \textbf{Effectiveness and Flexibility of Audio Integration.} As shown in Table \ref{tab:ab_new_ada},  we compare our baseline model which trained using only visual branch (line 1) against our audio-integrated model (line 2), which demonstrates the effectiveness of the incorporation of audio modality. To further validate the effectiveness and flexibility of our framework, we incorporate our IAMF module (Section \ref{sec:IAMF}) as a plug-in across advanced models (lines 3-6). Results indicate that all models achieve improvements across all metrics, with particularly notable gains on the challenging R1@7 metric, demonstrating that our approach effectively extracts meaningful information from audio modalities.

\begin{table} [tb!]
\caption{Effectiveness of audio integration for video moment retrieval. “$\uparrow$” denotes performance improvement when audio modality is introduced. 
}
\vspace{-4pt}
\label{tab:ab_new_ada}
\renewcommand{\arraystretch}{1.2}
\centering 
\scalebox{0.8}{
\begin{tabular}{@{}clllllll @{}}
\toprule
\multirow{2}{*}{\textbf{Line ID}}  &
\multirow{2}{*}{\textbf{Method}}   &
\multicolumn{2}{c}{\textbf{Charades-STA}} &
\multicolumn{2}{c}{\textbf{ActivityNet Captions}} & 
\\
\cmidrule(r){3-4} \cmidrule(r){5-6}
 & & \textbf{R1@7} & \textbf{mIOU}  & \textbf{R1@7} & \textbf{mIOU} \\
\hline
\#1 & Baseline & 39.52 & 52.76 & 26.18 & 43.21 \\
\#2 & Ours & $44.23_{4.71\uparrow}$ & $55.62_{2.86\uparrow}$ & $29.47_{3.29\uparrow}$ & $45.19_{1.98\uparrow}$ \\
\hline
\#3 & EMB \cite{huang2022video} & 39.25 & 53.09 & 26.07 & 45.59 \\
\#4 & EMB + Ours  & $43.15_{3.90\uparrow}$  & $54.53_{1.44\uparrow}$ & $28.44_{2.37\uparrow}$  & $46.69_{1.10\uparrow}$ \\
\hline
\#5 & EAMAT \cite{yang2022entity} & 41.96 & 54.45 & 25.77 & 42.19 \\
\#6 & EAMAT + Ours  & $44.08_{2.12\uparrow}$ & $55.59_{1.14\uparrow}$ & $27.38_{1.61\uparrow}$ & $43.27_{1.08\uparrow}$  \\
\bottomrule
\end{tabular}
}
\end{table}

\begin{table*} [tb!]
\caption{Comparison with audio-incorporated methods on Charades-STA and ActivityNet Captions. 
We use I3D \cite{carreira2017quo} as vision backbone with GloVe \cite{pennington2014glove} embeddings.
}
\label{tab:sota-ch&anet-i3d}
\renewcommand{\arraystretch}{1.2}
\centering 
\scalebox{0.8}{
\begin{tabular}{@{}lclllllllll}
\toprule
\multirow{2}{*}{\textbf{Method}} &
\multirow{2}{*}{\textbf{Audio}}   &
\multicolumn{4}{c}{\textbf{Charades-STA}} &
\multicolumn{4}{c}{\textbf{ActivityNet Captions}}   
\\
\cmidrule(r){3-6} \cmidrule(r){7-10}
& & \textbf{R1@3} & \textbf{R1@5}  & \textbf{R1@7} & \textbf{mIOU} & \textbf{R1@3} & \textbf{R1@5} & \textbf{R1@7} & \textbf{mIOU} \\
\hline
UMT \cite{liu2022umt} & \checkmark &  - & 48.31 & 29.25 & - & - & - & - & - \\
PMI-LOC w/o audio \cite{chen2020learning} & - & 56.84 & 41.29 & 20.11 & - & 60.16 & 39.16 & 18.02 & - \\
PMI-LOC \cite{chen2020learning} & \checkmark & $58.08_{1.24\uparrow}$ & $42.63_{1.34\uparrow}$ & $21.32_{1.21\uparrow}$ & - & $61.22_{1.06\uparrow}$ & $40.07_{0.91\uparrow}$ & $18.29_{0.27\uparrow}$ & - \\
QD-DETR w/o audio \cite{moon2023query} & - & -& 52.77 & 31.13 & - & - & - & - & - \\
QD-DETR \cite{moon2023query} & \checkmark & - & $55.51_{2.74\uparrow}$ & $34.17_{3.04\uparrow}$ & - & - & - & - & - & - \\
ADPN w/o audio \cite{chen2023curriculum} & - &  70.35  &  55.32  &  37.47 &  51.13  & 55.72  &  39.56  &  25.20 & 41.55 
\\
ADPN \cite{chen2023curriculum} & \checkmark &  ${71.99}_{1.64\uparrow}$  &  ${57.69}_{2.37\uparrow}$    &  ${41.10}_{3.63\uparrow}$ &  ${52.86}_{1.73\uparrow}$  & $57.16_{1.44\uparrow}$  &  $41.40_{1.84\uparrow}$  &  $26.31_{1.11\uparrow}$ & $42.31_{0.76\uparrow}$ 
\\
\hline
\textbf{IMG w/o audio} & -  & 72.37 & 56.34 & 39.52 & 52.76 & 59.19 & 41.51 & 26.18 & 43.21 \\
\textbf{IMG}  & \checkmark  & $\textbf{75.18}_{2.81\uparrow}$ & $\textbf{61.85}_{5.51\uparrow}$ & $\textbf{44.23}_{4.71\uparrow}$ & $\textbf{55.62}_{2.86\uparrow}$ & $\textbf{61.50}_{2.31\uparrow}$ & $\textbf{45.06}_{3.55\uparrow}$ & $\textbf{29.47}_{3.29\uparrow}$ & $\textbf{45.19}_{1.98\uparrow}$ \\
\bottomrule
\end{tabular} 
}
\vspace{-5px}
\end{table*}
\begin{table} [tb!]
\caption{Comparison with state-of-the-art methods on Charades-STA. We compare methods of using visual language models as backbone. “CLIP+SF” refers to SlowFast \cite{feichtenhofer2019slowfast} combined with CLIP \cite{radford2021learning} , “IV2” denotes InternVideo2 \cite{wang2024internvideo2}.
}
\vspace{-4px}
\label{tab:sota-ch&anet-intern}
\renewcommand{\arraystretch}{1.2}
\centering 
\scalebox{0.8}{
\begin{tabular}{@{}lcllllllll}
\toprule
\textbf{Method}   &
\textbf{backbone} & \textbf{R1@3} & \textbf{R1@5}  & \textbf{R1@7} & \textbf{mIOU} \\
\hline
UnLoc-L \cite{yan2023unloc} & CLIP & - & \textbf{60.80} & 38.40  & -  \\
Moment-DETR \cite{lei2021detecting} & CLIP+SF & - & 55.65 & 34.17 & -\\
BAM-DETR \cite{lee2024bam} & CLIP+SF & 72.93 & 59.95 &   39.38 &   52.33 \\
QD-DETR \cite{moon2023query}	& CLIP+SF & - & 57.31 &	32.55  &	-  \\
TR-DETR \cite{sun2024tr} & CLIP+SF & - &	57.61 &	33.52 &	-\\
UniVTG \cite{lin2023univtg} & CLIP+SF & 70.81 &	58.01 &	35.65 &	50.10 \\
FlashVTG \cite{cao2024flashvtg} & CLIP+SF & - & 60.11 & 38.01 & - \\
\hline
\textbf{IMG w/o audio} & CLIP+SF & 70.25 & 54.12 & 37.72 & 51.65\\
\textbf{IMG} & CLIP+SF &  $\textbf{74.44}_{3.38\uparrow}$ & $59.76_{5.64\uparrow}$ & $\textbf{42.93}_{5.21\uparrow}$	& $\textbf{55.03}_{3.38\uparrow}$ \\
\bottomrule
InternVideo2 \cite{wang2024internvideo2} & IV2 &79.70 & 70.03 & 48.95 & 58.79  \\
FlashVTG \cite{cao2024flashvtg} & IV2  & - & 70.32 & 49.87 & -   \\
SG-DETR \cite{gordeev2024saliency} & IV2 & - & 70.20 & 49.50 & 59.10 \\
\hline
\textbf{IMG w/o audio} & IV2  &  78.58 & 66.08 & 48.69	& 58.46 \\
\textbf{IMG} & IV2  & $\textbf{82.02}_{3.44\uparrow}$	& $\textbf{70.81}_{4.73\uparrow}$	& $\textbf{54.33}_{5.64\uparrow}$	& $\textbf{62.25}_{3.79\uparrow}$ \\
\bottomrule
\end{tabular} 
}
\vspace{-2px}
\end{table}

\vspace{-3px}
\subsection{Performance Comparison}
\vspace{-2px}
In Table \ref{tab:sota-ch&anet-i3d}, we evaluate our IMG on Charades-STA and ActivityNet Captions and compare it with existing audio-incorporated VMR methods. Furthermore, we list the results for the audio-incorporated method when trained without audio.
On Charades-STA and ActivityNet Captions, our IMG achieves the best performance on all metrics. 
By comparing with the visual branch, we find that introducing audio greatly improves sample training which demonstrates that audio modality can play an important role in assisting VMR. Moreover, our proposed methodology achieves markedly superior performance gains compared to existing approaches, which substantiates our methodological advantage.

In Table \ref{tab:sota-ch&anet-intern}, we evaluate IMG on Charades-STA, comparing it with state-of-the-art VMR methods that employ visual language models as backbones. IMG with InternVideo2 \cite{wang2024internvideo2} achieves the best performance across all metrics when the audio modality is incorporated. This highlights not only the generalization strength of our method under strong backbone settings, but also the critical role of audio cues in improving retrieval performance.


\begin{table} [tb!]
\caption{Performance comparison on Charades-AudioMatter. All methods utilize I3D \cite{carreira2017quo} backbone.}
\vspace{-4px}
\label{tab:sota-cam}
\renewcommand{\arraystretch}{1.2}
\centering 
\scalebox{0.8}{
\begin{tabular}{lcllll}
\toprule
\textbf{Method}  & \textbf{Audio} & \textbf{R1@3} & \textbf{R1@5} & \textbf{R1@7} & \textbf{mIOU} \\
\hline
SeqPAN \cite{zhang2021parallel} & - & 79.30 & 67.17 & 48.96 & 58.74\\
EAMAT \cite{yang2022entity} & - & 78.30 & 68.25 & 48.88 & 58.90\\
EMB \cite{huang2022video} & - & 77.81 & 67.00 &  47.96 & 58.66\\
ADPN w/o audio \cite{chen2023curriculum} & - & 77.89 & 64.42 & 44.64 & 56.98\\
ADPN \cite{chen2023curriculum} & \checkmark & $78.65_{0.76\uparrow}$ & $66.75_{2.33\uparrow}$ & $49.71_{5.07\uparrow}$ & $59.85_{2.87\uparrow}$ \\
\hline
\textbf{IMG w/o audio} & - & 77.89 & 65.92 & 47.58 & 58.35\\
\textbf{IMG} & \checkmark & $\textbf{82.74}_{4.85\uparrow}$ & $\textbf{71.93}_{6.01\uparrow}$ & $\textbf{54.27}_{6.69\uparrow}$ & $\textbf{62.76}_{4.41\uparrow}$\\
\bottomrule
\end{tabular}
}
\vspace{-2px}
\end{table}
 To highlight that IMG effectively mines audio modal information, we conducted experiments on Charades-AudioMatter where the audio data are more consistent and reliable. We compare our method with open-source methods that exhibit competitive performance on Charades-STA. As presented in Table \ref{tab:sota-cam}, our IMG achieves state-of-the-art performance, particularly on R1@7, thereby establishing a substantial lead over all other compared models. This result underscores the effectiveness of our model in extracting and utilizing audio modality and compare to ADPN, IMG exhibits superior proficiency in the integration of audio.

\vspace{-3px}
\subsection{Qualitative Analysis}
\vspace{-2px}

\begin{figure}[tb!]
  \centering
   \includegraphics[width=1\linewidth]{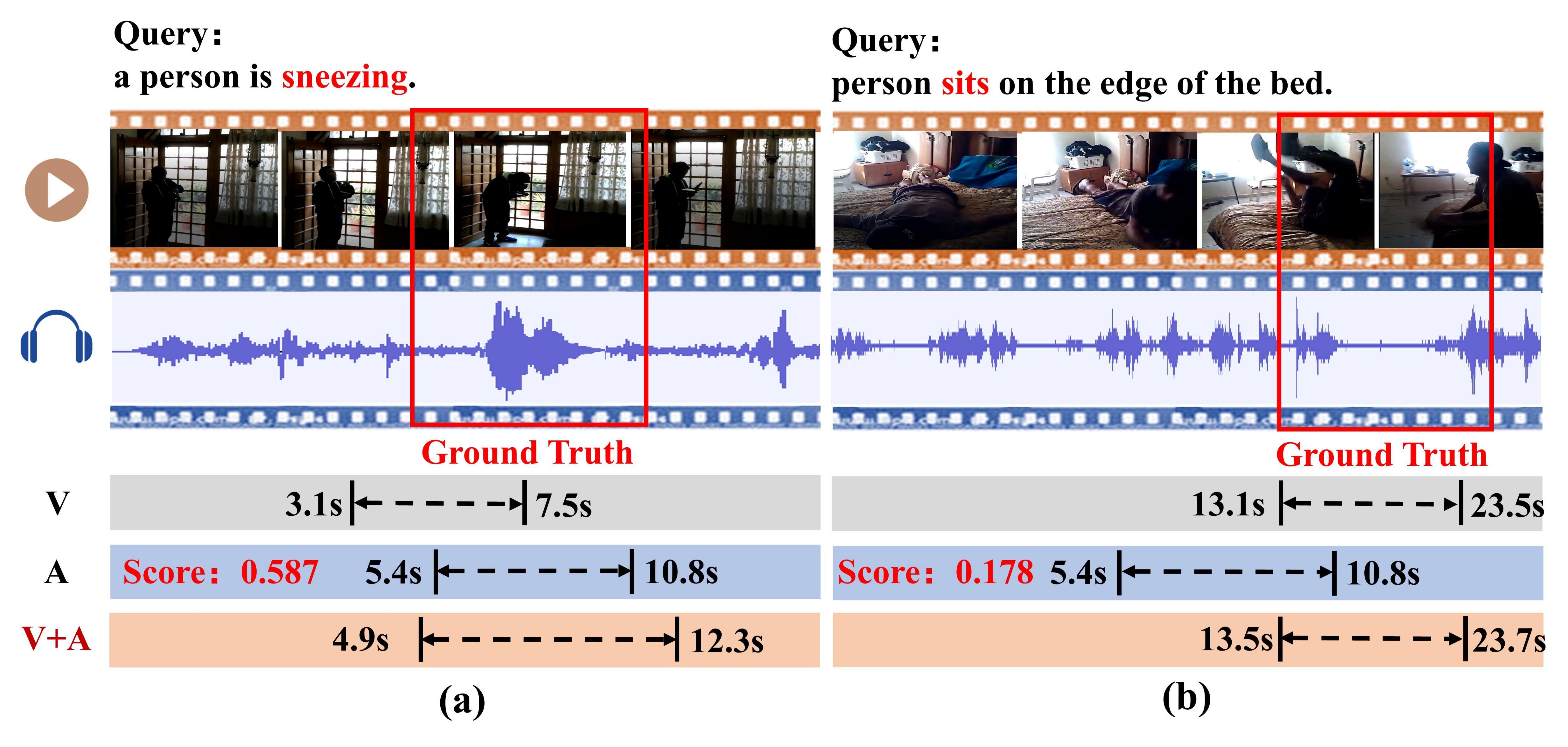}
   \vspace{-20pt}
   \caption{Two samples were selected from Charades-STA.}
   \vspace{-15pt}
   \label{fig:quantity1}
\end{figure}

As shown in Figure \ref{fig:quantity1}(a), we observe that the action "sneeze" is not clearly visible, leading to inaccurate predictions. In contrast, the audio prominently captures the action, with an AIP-predicted importance of 0.587, which helps correct the error in the fusion branch. In Figure \ref{fig:quantity1}(b), the action "sits" lacks distinct acoustic semantics, causing inaccurate inferences in the audio branch. AIP assigns an importance score of only 0.178, thereby reducing the fusion branch's reliance on audio.

\begin{figure}[t!]
  \centering
   \includegraphics[width=1\linewidth]{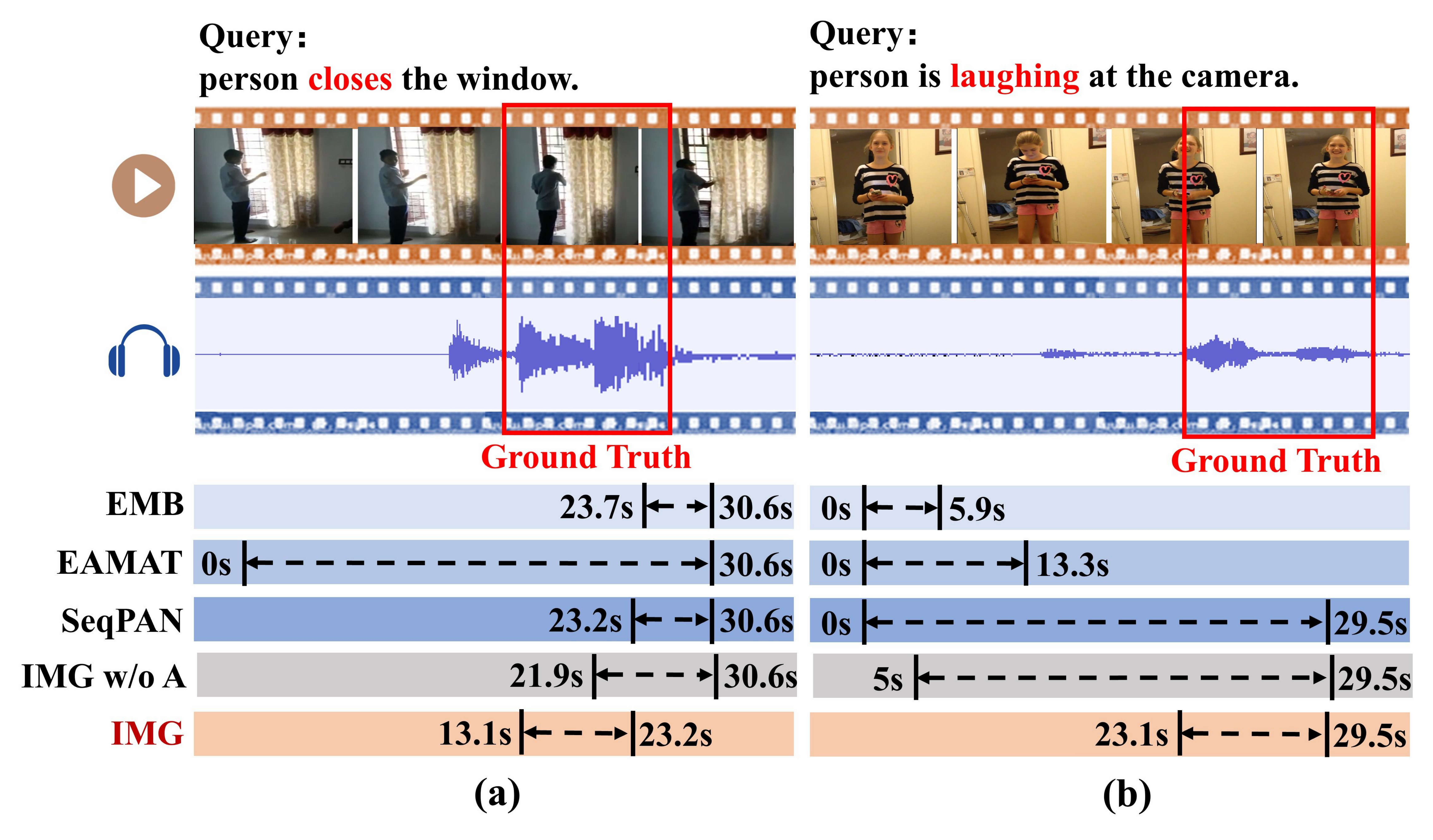} 
   \vspace{-20pt}
   \caption{Two samples were selected from Charades-AudioMatter. (a) appeared an occlusion interfering with the visual field, while (b) depicted a visually insignificant action.}
   \label{fig:supply_quan}
   \vspace{-15px}
\end{figure}

We also perform qualitative analysis on Charades-AudioMatter, comparing with methods that do not introduce audio. For Figure ~\ref{fig:supply_quan}(a), the window is partially obscured by curtains, which significantly increases the difficulty of visual-only retrieval for the action "closes the window". EMB, EAMAT, SeqPAN, and IMG without audio failed to retrieve accurately as they relied solely on vision. In contrast, IMG leveraged the acoustic semantics, allowing for more accurate retrieval. For Figure ~\ref{fig:supply_quan}(b), visual-only retrieval is also challenging due to the subtle movements associated with "laugh" and minimal scene variation. However, the prominent acoustic signal of "laugh" enabled IMG to effectively pinpoint the corresponding timestamp.

\vspace{-4px}
\section{Conclusion}
\vspace{-2px}
In this paper, we propose a novel Importance-aware Multi-Granularity fusion model (IMG) to handle the flexible audio-vision-text reasoning for the VMR task. To explore the audio's uncertainty, we propose an audio importance predictor to utilize the retrieval loss of the model to generate dynamic pseudo-labels for supervision and dynamically assign weights to different samples of audio to provide better audio-context guidance. We also propose a multi-granularity visual-audio fusion network to fully fuse audio and visual modality from local- to event- and global-level for complementary learning. A new dataset Charades-AudioMatter is further introduced to validate the model's capability in integrating audio modality. Experiments have proven the effectiveness of our proposed approach.

\begin{acks}
This work was supported by the Pioneer and Leading Goose R\&D Program of Zhejiang (No. 2024C01110), National Natural Science Foundation of China (No. 62472385), Young Elite Scientists Sponsorship Program by China Association for Science and Technology (No. 2022QNRC001), Public Welfare Technology Research Project of Zhejiang Province (No. LGF21F020010), Fundamental Research Funds for the Provincial Universities of Zhejiang (No. FR2402ZD) and Zhejiang Provincial High-Level Talent Special Support Program.
\end{acks}

\balance
\bibliographystyle{ACM-Reference-Format}
\bibliography{sample-base}
\clearpage
\appendix
We report more technical details and more experimental results which are not included in the paper due to space limit:

\begin{itemize}
    \item Detailed analysis of dataset Charades-AudioMatter including:
    \begin{itemize}
        \item Dataset construction (Section ~\ref{sec:data_cons}).
        \item Statistical analysis  (Section ~\ref{sec:stat_anal}).
    \end{itemize}
    \item Experiments on ActivityNet Captions including:
    \begin{itemize}
        \item Ablation studies on fusion strategies (Section ~\ref{sec: fs}).
        \item Ablation studies on additional model structures (Section ~\ref{sec: ams}).
        \item Qualitative analysis 
        (Section ~\ref{sec: qa}).
    \end{itemize}
    \item Additional experiments including:
    \begin{itemize}
        \item Experiments on hyperparameters (Section ~\ref{sec:hyp}) including threshold $\epsilon_{min}$, temperature $\gamma$ and others.
        \item Experiments on efficiency (Section ~\ref{sec:flop}), Event-Level Fusion module (Section ~\ref{sec:slot}), weak supervision (Section ~\ref{sec:weak}), failed AIP (Section ~\ref{sec:failed_aip}) and audio importance distribution (Section ~\ref{sec:aid}).
    \end{itemize}
    \item Implement details (Section ~\ref{sec:im_de}).
\end{itemize}

\section{Charades-AudioMatter Dataset Construction}
\label{sec: statistical analysis}
\subsection{Dataset Construction}
\label{sec:data_cons}
In this section, we introduce the dataset Charades-AudioMatter in detail. To ensure the high quality of the Charades-AudioMatter dataset and the reliability of experimental results, the dataset construction underwent a rigorous screening process. The dataset was annotated by six postgraduate students with experience in multimodal learning. Each instance was independently labeled by two annotators, with disagreements adjudicated by a third annotator. The validity and relevance of the audio data were carefully evaluated through the following processes: 

\noindent \textbf{Validity of the Audio.} Given a sample,  the audio modality is first subjected to a validity assessment. Samples containing significant background noise or lacking any sound were excluded, as such audio lacks meaningful information and cannot contribute effectively to VMR. This process is employed for rapid preliminary screening.

\noindent \textbf{Correlation between Audio and Query Text.} After the initial screening, each sample was manually evaluated through a combination of audio and visual to determine whether the query text was dependent on the audio. For instance, query describing static actions (\textit{e.g.}, “sitting,” “looking,” “standing”) were almost excluded because the audio does not provide meaningful cues for these actions. Similarly, for actions typically associated with audio cues (\textit{e.g.}, “laughing,” “closing the door”), if the audio in a specific instance lacked sound or the sound was too faint, the sample was marked as invalid and excluded. This step ensured the relevance of audio to the text by integrating auditory judgment with semantic analysis of the text.

\begin{figure}[t!]
  \centering
   \includegraphics[width=1\linewidth]{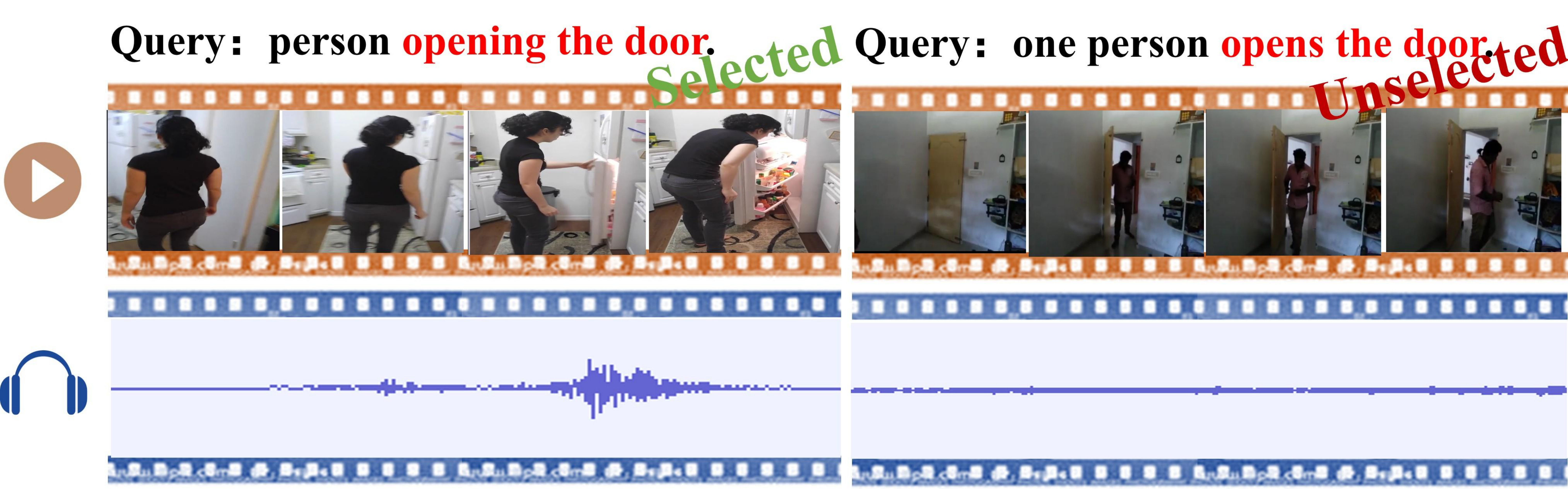} 
   \caption{For the same activity of ``open the door", we reviewed and listened specific samples, ultimately choosing the left one where the sound is clearly communicated, discarding the other where there is almost no corresponding sound.}
   \label{fig:supply_sample}
\end{figure}
\begin{table} [htbp]
\renewcommand{\arraystretch}{1.2}
\caption{Statistical analysis of the dataset Charades-AudioMatter. We compare the selected activity categories with the unselected one.
}
\vspace{-4pt}
\label{tab:anal}
\centering 
\scalebox{0.8}{
\begin{tabular}{lc|lc}
\toprule
\textbf{\textit{Selected} Activity} & \textbf{Count} & \textbf{\textit{Unselected} Activity} & \textbf{Count}\\
\hline
open (door/cabinet/...) & 241 & sit (on bed/chair/...) & 218
\\
close (door/closet/...)   & 150 & hold & 147
\\
put (bag/grpceries/...) & 138 & (un)dress & 111 
\\
run & 90 & look & 85
\\
turn on/off (light/tv/...) & 89 & stand & 59
\\
throw (broom/shoes/…)  &  66 & smile & 55
\\
take (vacuum/food/...) & 56 & watch & 48
\\
laugh &52 & read & 32
\\
eat & 41 &  awake & 38
\\
 wash (hand/glass/...) &29 & take a picture & 30
\\
drink& 28  & play (phone/camera/...) & 23
\\
walk & 25 & snuggle with (pillow/...) & 20
\\
cook & 22  & (fix/adjust) hair & 19 
\\
pour (water/coffee/...) & 16 & lay & 18 
\\
sit down & 13
\\
talk & 10
\\
\bottomrule
\end{tabular}
}
\end{table}
\noindent \textbf{Temporal Alignment of Audio and Video.}
After the screening steps above, we evaluate the temporal alignment between visual and audio modalities. Specifically, manual timestamp annotation was performed solely based on the audio and query text, followed by IoU computation with the ground truth. Samples exhibiting an IoU score below 0.3 are discarded. This process ensures the validation of temporal consistency between audio and video modalities, effectively filtering out severely misaligned samples (e.g., those with significant audio delays or excessive offsets).

Upon completing the labeling process, a random sampling procedure was conducted to evaluate the reliability and consistency of the annotations, and the final inter-annotator agreement exceeded 95\%. This rigorous, multi-step approach ensures that the dataset adheres to high-quality standards while providing a robust foundation for advancing research in the VMR task.

\subsection{Statistical Analysis}
\label{sec:stat_anal}
To further demonstrate the effectiveness of our proposed dataset Charades-AudioMatter, we conduct activity category-wise analysis in Table ~\ref{tab:anal}. We sorted the categories of selected activities and compared them with other unselected activities. As we can see from the table, selected activities tend to have more significant differentiating sounds such as ``open", ``put", and ``run", while unselected activities do not tend to convey sounds such as ``sit on", ``hold", and ``look". But we can't exactly classify audio validity by activity, we give samples in Figure ~\ref{fig:supply_sample}.

\begin{figure}[t]
  \centering
    {\includegraphics[width=0.48\linewidth]{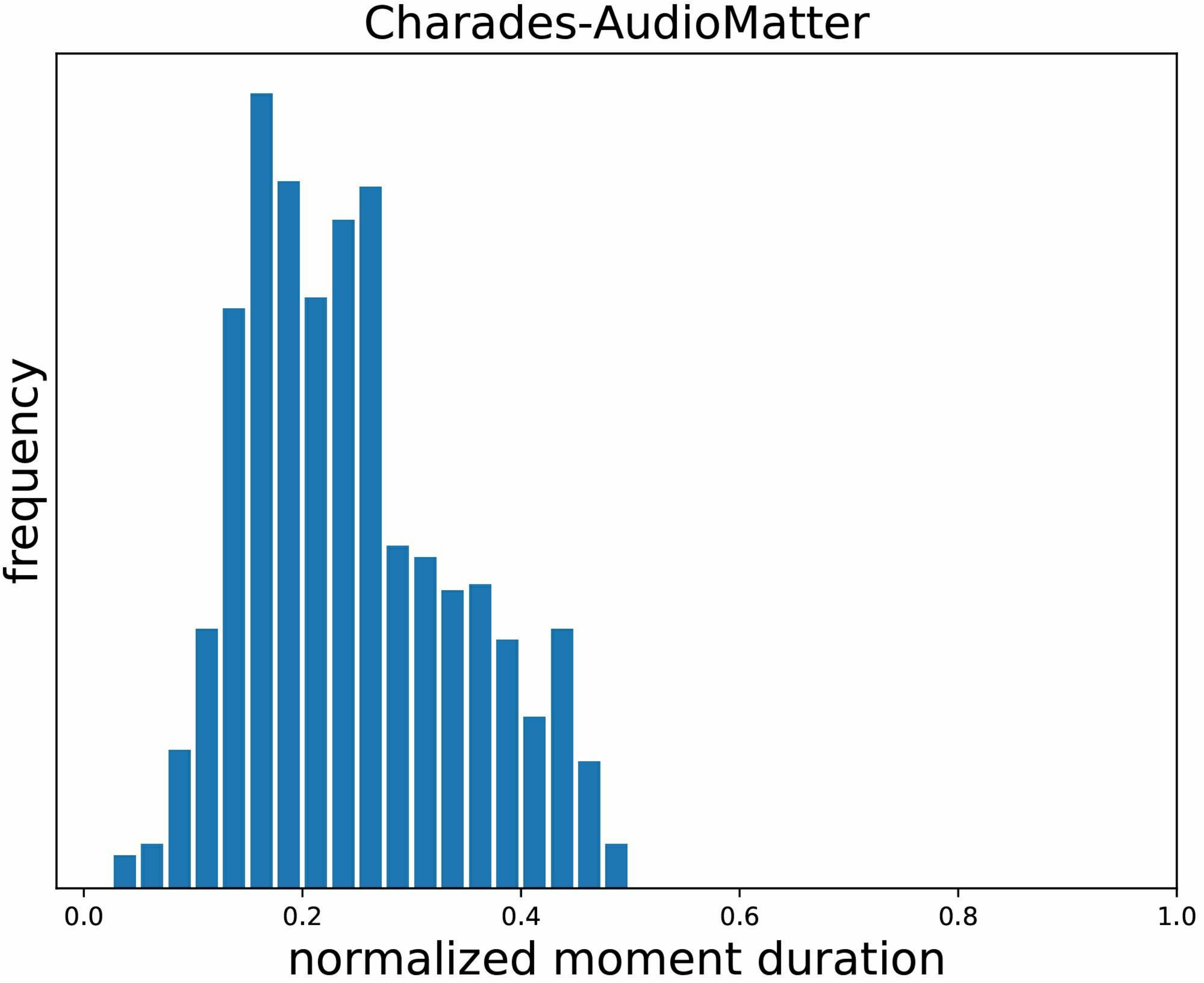}}
    {\includegraphics[width=0.48\linewidth]{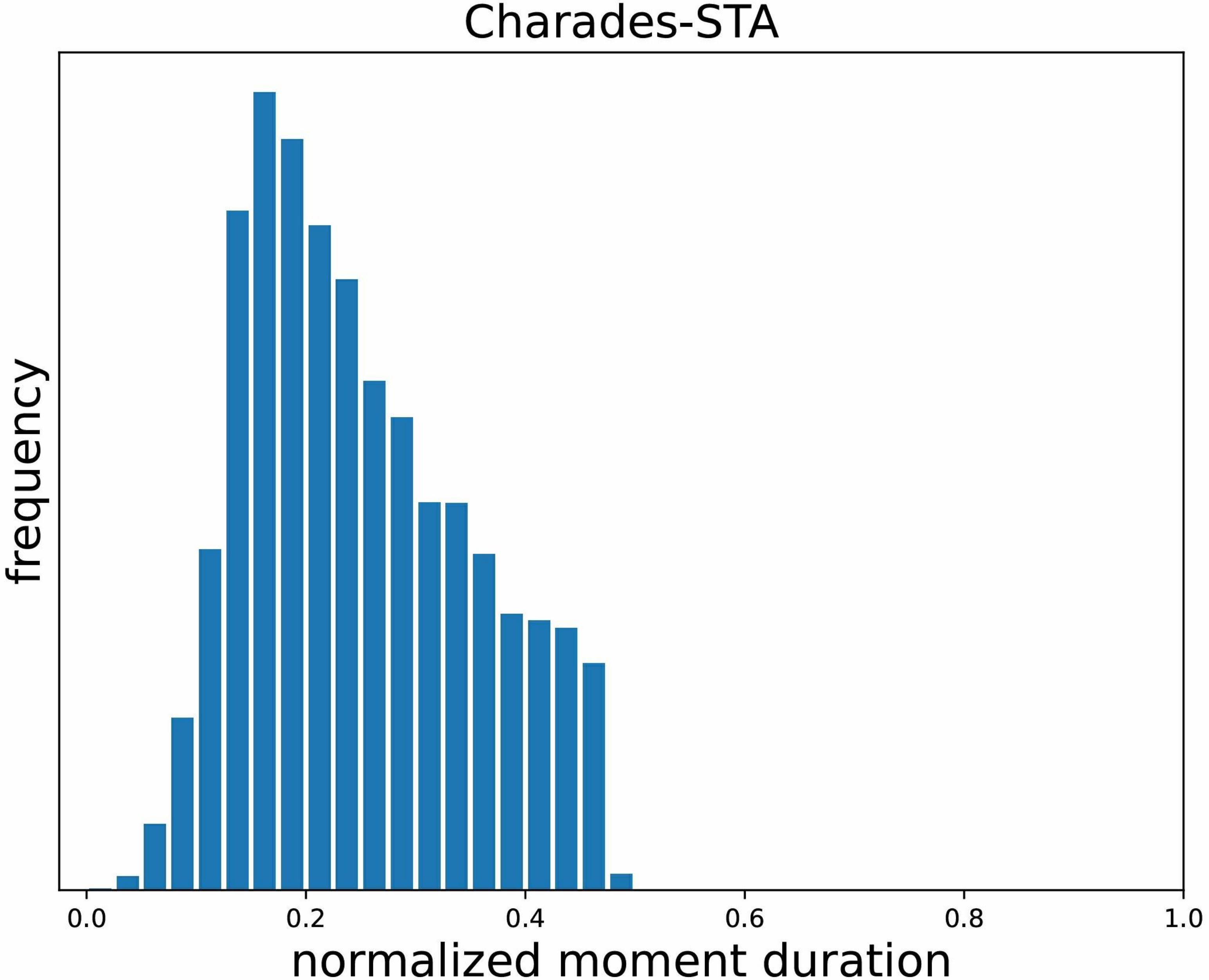}}
  \caption{Comparison between Charades-AudioMatter and Charades-STA in terms of moment duration.}
  \label{fig:supply_time}
\end{figure}

Table \ref{fig:supply_time} shows the frequency distribution of normalized moment durations in Charades-AudioMatter compared with the original Charades-STA. Our purposed Charades-AudioMatter maintains comparable diversity and roughly follows the original Charades-STA in duration distribution, which validates the rationality of our proposed dataset.

\section{Experiments on ActivityNet Captions}
\label{sec:ab_anet}

To further verify the general effectiveness of the crucial contributions in our proposed IMG, we conduct more experiments on ActivityNet Captions.
\subsection{Ablation studies on fusion strategies}
\label{sec: fs}
We verify the effectiveness of fusion strategies on ActivityNet Captions. As shown in Table ~\ref{tab:ab_anet}, each of the proposed fusion strategies consistently yields performance improvements, underscoring their efficacy. These results also demonstrate that integrating features of varying granularities provides complementary benefits, leading to superior overall performance.

\begin{table} [tb!]
\renewcommand{\arraystretch}{1.2}
\centering 
\caption{Ablation studies of fusion strategies on ActivityNet Captions.}
\scalebox{0.8}{
\begin{tabular}{ccccccc}
\toprule
 \textbf{Local} & \textbf{Event} & \textbf{Global} & \textbf{R1@3} & \textbf{R1@5} & \textbf{R1@7} & \textbf{mIOU} \\
\hline
 \checkmark & - & - & 60.08 & 43.70 & 27.66 & 43.96 \\
 - & \checkmark & - & 59.13 & 42.68 & 26.83 & 43.54 \\
 - & - & \checkmark & 58.57 & 42.12 & 27.00 & 43.22\\
\hline
 \checkmark & \checkmark & - & 59.21 & 43.60 & 28.71 & 44.17 \\
 \checkmark & - & \checkmark & 61.25 & 43.90 & 28.92 & 44.78 \\
 - & \checkmark & \checkmark & 59.84 & 43.69 & 28.19 & 44.05 \\
\hline
 \checkmark & \checkmark & \checkmark & \textbf{61.50} & \textbf{45.06} & \textbf{29.47} & \textbf{45.19}  \\

\bottomrule
\end{tabular}
}
\label{tab:ab_anet}
\end{table}
\begin{table} [tb!]
\renewcommand{\arraystretch}{1.2}
\centering
\caption{Ablation studies of each component on ActivityNet Captions.}
\scalebox{0.8}{
\begin{tabular}{lcccc}
\toprule
\textbf{Method} & \textbf{R1@3} & \textbf{R1@5} & \textbf{R1@7} & \textbf{mIOU} \\
\hline
IMG w/o AIP & 59.81	& 43.40	 & 28.06 & 44.49
\\
IMG w/o pseudo-label & 58.10 & 42.00 & 27.76 & 43.52
\\
IMG w/o CKD  &  59.95 & 43.89 & 28.49 & 44.47
\\
IMG & \textbf{61.50} & \textbf{45.06} & \textbf{29.47} & \textbf{45.19} \\
\bottomrule
\end{tabular}
}
\label{tab:s_ab_mo}
\end{table}
\subsection{Ablation studies on additional model structures}
\label{sec: ams}
We also verify the effectiveness of additional model structures on ActivityNet Captions. As presented in Table ~\ref{tab:s_ab_mo}, lines 1 and 2 illustrate the performance of the model without the audio importance predictor and the pseudo-label constraint, respectively. These results indicate that the proposed pseudo-label mechanism enhances decision-making within the audio importance predictor and ultimately improves performance. Finally, line 3 quantifies the effect of ablating cross-modal knowledge distillation, further validating the contributions of this component to the overall framework.

\subsection{Qualitative analysis}
\label{sec: qa}
\begin{figure}[tb!]
  \centering
   \includegraphics[width=1\linewidth]{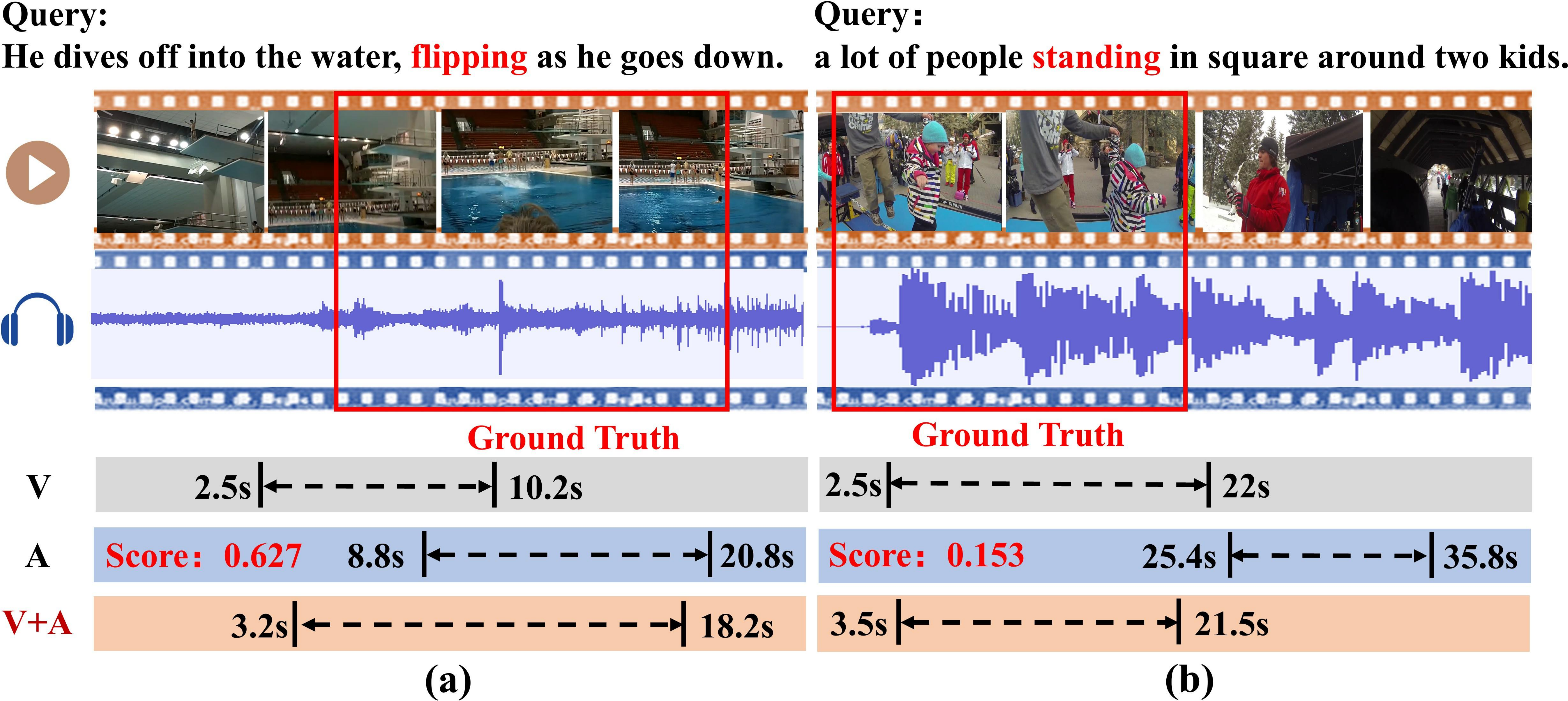} 
   \caption{Two samples were selected from ActivityNet Captions.}
   \label{fig:quantity2}
\end{figure}
In order to demonstrate our proposed IMG more intuitively, we have selected examples on ActivityNet Captions for visual presentation. As shown in Figure \ref{fig:quantity2}(a) "dives" is visible in the frames, but "flipping" is not distinctly captured. However, both actions exhibit clear acoustic semantics, allowing the fusion branch to make a more accurate prediction. In Figure \ref{fig:quantity2}(b), the visual presentation is highly noticeable, while the audio consists entirely of background music, and as a result, the fusion branch is not misled by the audio.

\begin{figure}[t]
  \centering
    \subfigure[Performance curves at different thresholds.]{\includegraphics[width=0.48\linewidth]{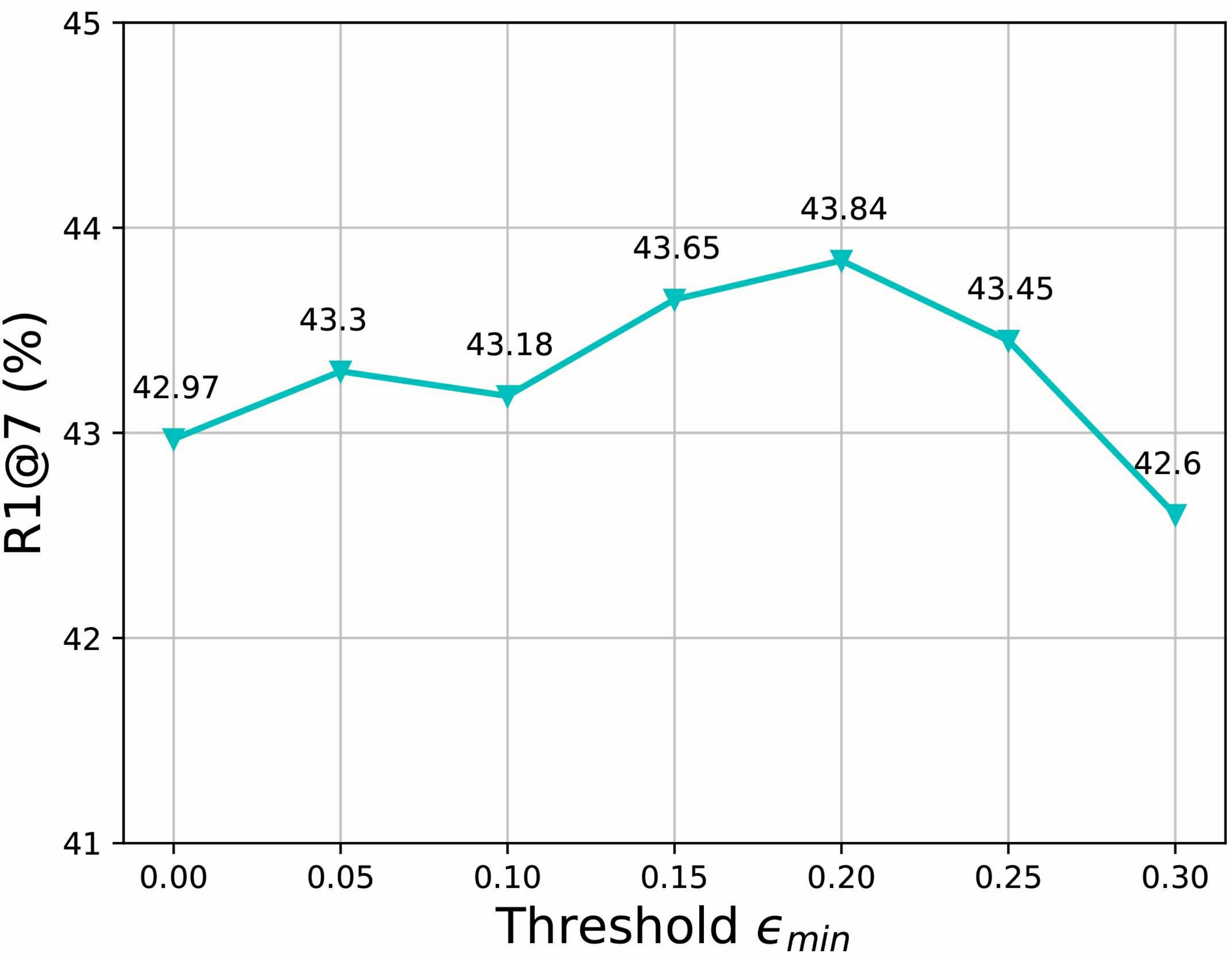}}
    \subfigure[Performance curves at different temperatures.]{\includegraphics[width=0.48\linewidth]{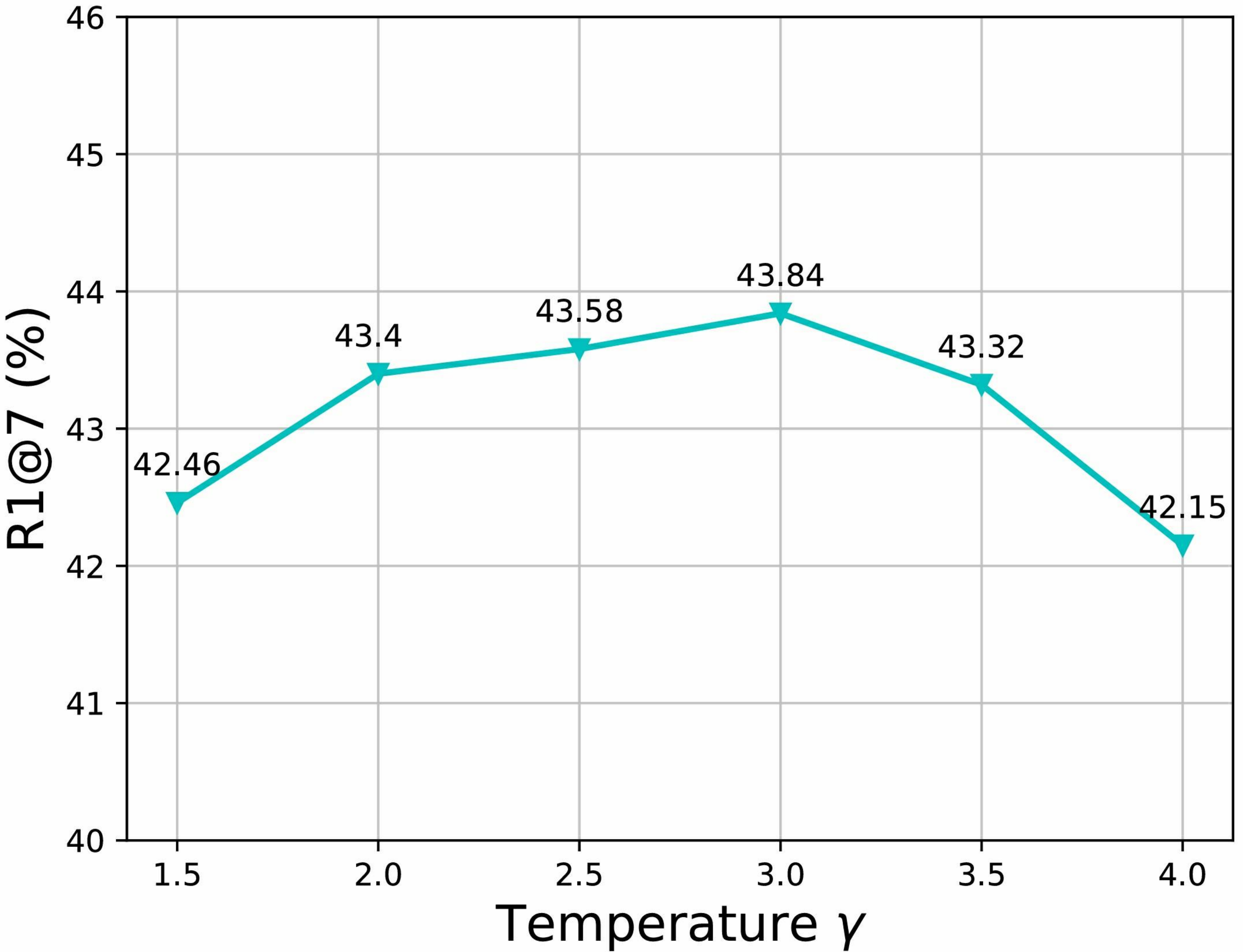}}
  \caption{Experiments with different hyperparameters, (a) threshold $\epsilon_{min}$, (b) temperature $\gamma$.}
  \vspace{-8pt}
  \label{fig:ab_hyp}
\end{figure}

\subsection{Ablation studies on additional model structures}
\label{sec: ams}
We also verify the effectiveness of additional model structures on ActivityNet Captions. As presented in Table ~\ref{tab:s_ab_mo}, lines 1 and 2 illustrate the performance of the model without the audio importance predictor and the pseudo-label constraint, respectively. These results indicate that the proposed pseudo-label mechanism enhances decision-making within the audio importance predictor and ultimately improves performance. Finally, line 3 quantifies the effect of ablating cross-modal knowledge distillation, further validating the contributions of this component to the overall framework.

\section{Additional experiments}
\subsection{Experiments on hyperparameters}
\label{sec:hyp}
We conduct ablation studies on two critical hyperparameters, threshold $\epsilon_{min}$ and temperature $\gamma$. As detailed in Figure~\ref{fig:ab_hyp}, our analysis reveals that selecting an optimal threshold and temperature significantly enhances the model’s ability to learn the relative importance of visual and audio features, thereby improving overall performance. Conversely, setting the threshold too low may cause the model to mistakenly assign importance to noisy semantic features, while setting it too high can lead the model to disregard valuable samples containing relevant semantic information. Similarly, an excessively low temperature coefficient results in overly rigid decision-making by the model, whereas an excessively high coefficient diminishes the model’s sensitivity to the two feature types, ultimately impairing performance.

In Table \ref{tab:tau}, we conduct an ablation study on the temperature coefficient $\tau$ in cross-modal knowledge distillation module. The results indicate that our method is relatively insensitive to $\tau$.
\begin{table} [tb!]
\renewcommand{\arraystretch}{1.2}
\centering
\caption{Ablation studies on hyperparameters $\tau$ on visual branch.}
\scalebox{0.8}{
\begin{tabular}{lcccc}
\toprule
\textbf{$\tau$} & \textbf{0.5} & \textbf{1.0} & \textbf{2.0} & \textbf{4.0} \\
\hline
R1@7 & 42.10 &	42.61 &	43.44 &	42.90
\\
\bottomrule
\end{tabular}
}
\label{tab:tau}
\end{table}

For loss-related parameters, both $\lambda_{1}$ and $\lambda_{2}$ are crucial to the model. We set $\lambda_{1}$ = 5 and $\lambda_{2}$ = 10 to balance the respective loss terms, ensuring they are on a similar scale. $\lambda_{3}$ controls the auxiliary loss, and we set $\lambda_{3}$ = 0.5, significantly smaller than the others. These values were determined based on grid search and empirical validation. 
 
\subsection{Experiments on efficiency}
\label{sec:flop}
As shown in the Table \ref{tab:flops}, we evaluate the efficiency of our proposed method by measuring both the FLOPs and the number of parameters during inference. Compared to several open-source methods, our model maintains low computational and parameter overhead while achieving better performance. Additionally, we also report the results of key modules in our method including Audio Importance Predictor (AIP) ,Multi-Granularity Fusion (MGF) .
\begin{table} [tb!]
\renewcommand{\arraystretch}{1.2}
\centering
\caption{Comparison on flops and params.}
\scalebox{0.8}{
\begin{tabular}{lccc}
\toprule
\textbf{Method} & \textbf{Flops(G)} & \textbf{Params(M)} & \textbf{R1@7}  \\
\hline
EAMAT&9.97&94.12&41.96 \\
BAM-DETR&1.39&13.43&39.38 \\
FlashVTG&1.05&8.73&38.01 \\
QD-DETR&0.82&6.36&32.55 \\
Moment-DETR&0.26&3.23&38.01 \\
ADPN&0.34&1.54&41.10 \\
\hline
\textbf{IMG}&0.38&3.31&\textbf{44.23} \\
  —AIP & $9.87 \times 10^{-5}$ & $5.12 \times 10^{-4}$ & - \\
  —MGF & 0.20 & 1.91 & - 
\\
\bottomrule
\end{tabular}
}
\label{tab:flops}
\end{table}

\subsection{Experiments on Event-Level Fusion}
\label{sec:slot}
 \textbf{Slots with supervision.} Since the moment boundary label is coarse-level and does not have its contained event split, we cannot utilize it to provide explicit event-level supervision. Therefore, we adopt the unsupervised slot attention mechanism to implicitly learn the potential event contexts. Specifically, we utilize the moment boundary label to additionally supervise the global content of all events. After slot interaction, the event-level sequences are globally projected into a 1D sequence via an MLP and Sigmoid for supervision with binary cross-entropy loss. As shown in Table \ref{tab:sup_slot}, the global supervision yields a similar performance compared to the unsupervised one. We assume that this is due to the limited granularity of available supervision in VMR, while the unsupervised version can implicitly learn the potential events.
\begin{table} [tb!]
\renewcommand{\arraystretch}{1.2}
\centering
\caption{Ablation studies on supervised slots.}
\scalebox{0.8}{
\begin{tabular}{lcccc}
\toprule
\textbf{Method} & \textbf{R1@3} & \textbf{R1@5} & \textbf{R1@7} & \textbf{mIOU} \\
\hline
slot w/ supervision & 73.77 & 59.60 & 41.19 & 54.51 \\
slot w/o supervision & 74.84 & 59.92 & 41.32 & 54.83

\\
\bottomrule
\end{tabular}
}
\label{tab:sup_slot}
\end{table}

\textbf{Number of slots and iter.} Table \ref{tab:num_iter} summarizes the performance with varying numbers of slots and iterations. While increasing the number of slots and iterations increases computational overhead, we find that using 2 or 3 slots with 3 or 4 iterations achieves satisfactory performance. Moreover, using 3 slots and 3 iterations offers a good balance between performance and computational cost.
\begin{table} [tb!]
\renewcommand{\arraystretch}{1.2}
\centering
\caption{Performance under different slot numbers and iterations.}
\scalebox{0.8}{
\begin{tabular}{cccccc}
\toprule
\textbf{\#Slot \textbackslash\ \#Iter} & \textbf{1} & \textbf{2} & \textbf{3} & \textbf{4} & \textbf{5} \\
\midrule
2 & 40.45 & 40.75 & 41.23 & 41.01 & 40.76 \\
3 & 40.23 & 40.92 & \textbf{41.32} & 41.20 & 41.25 \\
4 & 39.88 & 40.45 & 41.10 & 41.22 & 40.95 \\
5 & 39.20 & 39.70 & 40.12 & 40.70 & 40.32 \\
\bottomrule
\end{tabular}
}
\label{tab:num_iter}
\end{table}

\subsection{Experiments on weak supervision}
\label{sec:weak}
\noindent
To assess the robustness of our model under limited supervision, we further conduct evaluation using reduced training data (70\%, 80\%, and 90\% subsets). As shown in Table~\ref{tab:weak}, our model largely maintains its performance and consistently outperforms the strong baseline ADPN.
\begin{table}[tb!]
\renewcommand{\arraystretch}{1.2}
\centering
\caption{Performance under different training set sizes to evaluate weak supervision capability.}
\label{tab:weak}
\scalebox{0.8}{
\begin{tabular}{lccccc}
\toprule
\multirow{2}{*}{\textbf{Method}} & \multicolumn{4}{c}{\textbf{Samples for train (\%)}} \\
\cmidrule(lr){2-5}
& \textbf{70} & \textbf{80} & \textbf{90} & \textbf{100} \\
\midrule
ADPN & 37.07 & 38.78 & 39.47 & 41.10 \\
IMG  & 41.21 & 43.29 & 43.52 & 44.23 \\
\bottomrule
\end{tabular}
}
\end{table}

\subsection{Audio importance distribution}
\label{sec:aid}
\noindent
Table \ref{tab:dis} shows the distribution of audio importance scores, where most samples fall within the range of 0.15 to 0.45, further validating that audio serves as an auxiliary modality.
\begin{table} [tb!]
\renewcommand{\arraystretch}{1.2}
\centering
\caption{Distribution of audio importance scores across samples.}
\scalebox{0.8}{
\begin{tabular}{lccccc}
\toprule
\textbf{Score Range} & \textless 0.15 & 0.15--0.25 & 0.25--0.35 & 0.35--0.45 & \textgreater 0.45 \\
\midrule
\textbf{Count} & 26 & 963 & 1861 & 665 & 205 \\
\bottomrule
\end{tabular}
}
\label{tab:dis}
\end{table}

\subsection{Experiments on failed AIP }
\label{sec:failed_aip}
\noindent
 We explore the impact when AIP mistakenly predicts audio importance to zero ($p$ = 0) on Charades-AudioMatter. As shown in Table \ref{tab:failed_aip}, such change degrade the performance. The results not only show the importance of audio clues, but also demonstrate the effectivenesss of our proposed audio-aware desgin. 
\begin{table} [tb!]
\renewcommand{\arraystretch}{1.2}
\centering
\caption{Ablation studies on AIP with zero importance.}
\scalebox{0.8}{
\begin{tabular}{lcccc}
\toprule
\textbf{Method} & \textbf{R1@3} & \textbf{R1@5} & \textbf{R1@7} & \textbf{mIOU} \\
\hline
IMG & 82.74 & 71.93 & 54.27 & 62.76 \\
IMG ($p$=0) & 80.27 & 70.22 & 50.96 & 59.84

\\
\bottomrule
\end{tabular}
}
\label{tab:failed_aip}
\end{table}

\section{Implement Details} 
\label{sec:im_de}
For all datasets, we set the initial learning rate to 0.0005, and the maximum number of frames to 128. We use AdamW \cite{loshchilov2017fixing} for optimization and linear decay scheduling, and the maximum epoch number is 100 for all of them with batch size 16. We use I3D \cite{carreira2017quo} as pretrained visual features on all datasets. For the audio pretraining models, following previous work \cite{chen2023curriculum}, we utilized PANN \cite{kong2020panns}, pretrained on AudioSet \cite{gemmeke2017audio} dataset, and VGGish \cite{hershey2017cnn}, pretrained on YouTube-100M \cite{hershey2017cnn} dataset, for Charades-STA/Charades-AudioMatter and ActivityNet Caption, respectively. We initialize words with 300d GloVe \cite{pennington2014glove} embeddings. To further demonstrate the generalizability of our model, we also use InternVideo2 \cite{wang2024internvideo2} and LLaMA \cite{touvron2023llama} for visual and textual backbone. We set $\epsilon_{min}$ to 0.2, $\gamma$ to 3 for Charades-STA/Charades-AudioMatter, $\epsilon_{min}$ to 0.1 and  $\gamma$ to 2 for ActivityNet Captions. All experiments are implemented on a single NVIDIA 3090 GPU.









\end{document}